\documentclass[aps,prb,twocolumn,superscriptaddress]{revtex4-2}
\usepackage{graphicx}
\usepackage{amsmath,amssymb}
\usepackage{physics}
\usepackage{xcolor}

\newcommand{\br}{\mathbf{r}}
\newcommand{\bk}{\mathbf{k}}
\newcommand{\bp}{\mathbf{p}}
\newcommand{\bA}{\mathbf{A}}
\newcommand{\bB}{\mathbf{B}}

\newcommand{\ba}{\mathbf{a}}
\newcommand{\bg}{\mathbf{g}}
\newcommand{\bR}{\mathbf{R}}

\newcommand{\bq}{\mathbf{q}}

\usepackage{hyperref}
\hypersetup{colorlinks,
linkcolor={blue},
citecolor={blue},
urlcolor={purple}}  

\begin{document}
\title{Revisiting Bloch electrons in magnetic field: Hofstadter physics via hybrid Wannier states}
\author{Xiaoyu Wang}
\affiliation{National High Magnetic Field Lab, Tallahassee, FL 32310}
\author{Oskar Vafek}
\affiliation{National High Magnetic Field Lab, Tallahassee, FL 32310}
\affiliation{Department of Physics, Florida State University, Tallahassee, Florida 32306, USA}
\begin{abstract}
    We revisit the Hofstadter butterfly for a subset of topologically trivial Bloch bands arising from a continuum free electron Hamiltonian in a periodic lattice potential. We employ the recently developed procedure -- which was previously used to analyze the case of topologically non-trivial bands [\href{https://journals.aps.org/prb/abstract/10.1103/PhysRevB.106.L121111}{Phys. Rev. B \textbf{106}, L121111 (2022)}] -- to construct the finite field Hilbert space from the zero-field hybrid Wannier basis states. Such states are Bloch extended along one direction and exponentially localized along the other. The method is illustrated for square and triangular lattice potentials and is shown to reproduce all the main features of the Hofstadter spectrum obtained from a numerically exact  Landau level expansion method. 
    In the regime when magnetic length is much longer than the spatial extent of the hybrid Wannier state in the localized direction we recover the well known Harper equation. Because the method applies to both topologically trivial and non-trivial bands, it provides an alternative and efficient approach to moir\'e materials in magnetic field.  
\end{abstract}

\maketitle

\section{Introduction}
The discovery of superconductivity and correlated insulating states in the twisted bilayer graphene (TBG) \cite{Cao2018a,Cao2018b} has invigorated the study of various 2D moir\'e electronic materials \cite{Cao2020,Tang2020,Liu2020,Wang2020,Regan2020,Park2021,Hao2021,Kometter2022}. The moir\'e superlattice is generated by stacking 2D layered structures either with a small twist or via microscopic lattice mismatch. Despite the large moir\'e unit cell containing as many as $\sim 10,000$ atoms, the low energy physics is dominated by only a few isolated narrow Bloch bands formed due to the moir\'e superlattice potential, motivating theoretical studies that focus on these low energy degrees of freedom \cite{Yuan2018,Koshino2018,Kang2018,Po2018,Kang2019, TBGI2020,TBGII2020,TBGIII2020,TBGIV2020,TBGV2020,Balents2020,Song2022}.

The large moir\'e unit cell has also enabled the study of magnetic field effects in such systems in the regime with considerable fraction of one full magnetic flux quantum per moir\'e unit cell. Electronic interaction effects intermixed with strong magnetic fields have been studied in various moir\'e materials, revealing not only Landau level degeneracies indicative of the symmetry-breaking phases at zero magnetic field, but also novel field-induced insulating states that can carry finite Chern numbers \cite{dean_hofstadters_2013,Lu2019,Yankowitz2019,Sharpe2019,Nuckolls2020,Pierce2021,Saito2021,Wu2021,Finney2022,Yu2022}. These experimental results in turn motivate further theoretical studies of Hofstadter physics \cite{Hejazi2019,Yahui2019,Biao2021LL,Jonah2021,Parker2021b,Jonah2022}. 

Traditionally, Bloch electrons in magnetic field $\bB$ have been studied either by solving the continuum Hamiltonian $\hat{H}(\br,\hat{\bp}+\frac{e}{c}\bA)$, where the magnetic vector potential satisfies $\bB=\nabla\times\bA$, or via Peierls phase substitution of hopping amplitudes \cite{Peierls1933,Luttinger1951,Harper1955,Hofstadter1976}. The latter is justified if the subset of Bloch bands of interest at zero field is amenable to a tight-binding description i.e. there is no topological obstruction to Wannierization. In the first approach, one calculates the matrix elements of $\hat{H}$, including the periodic lattice potential, in the Hilbert space spanned by Landau level (LL) wavefunctions (obtained without the periodic lattice potential), and diagonalizes the resulting matrix. To achieve numerical convergence within an energy window $W$ an upper LL index cutoff $N_c\sim \lambda W/\hbar \omega_c$ is needed, where $\omega_c=eB/m_ec$ is the cyclotron frequency, $m_e$ is the bare electron mass, and $\lambda$ is a number that increases with the strength of the periodic lattice potential $V$. For example, Ref.~\cite{Hejazi2019} has pointed out that an upper LL cutoff of $N_c \sim 25\phi_0/\phi$ is needed to faithfully reproduce the narrow band Hofstadter spectrum of TBG where $\phi_0=hc/e$ is the magnetic flux quantum, and $\phi$ is the magnetic flux per moir\'e unit cell.  This is a computationally intensive procedure especially at low $\bB$ when $N_c$  becomes large. In the second approach, the magnetic field effects are accounted for via Peierls substitution, i.e. replacing the intersite hopping amplitude $t_{ij}$ with $t_{ij} \exp\left(-i \frac{e}{\hbar c}\int_{\br_i}^{\br_j}\mathrm{d}\br\cdot\bA(\br)\right)$. The Peierls substitution has been used extensively in the literature due to its simplicity in addressing Hofstadter physics. However, it is unclear how to generalize the Peierls substitution to a subset of Bloch bands where 2D exponentially localized and symmetric Wannier orbitals cannot be constructed \cite{Soluyanov2011,Po2018b,XW2020}. 

In an earlier work of ours \cite{XW2022}, to address the Hofstadter physics in TBG, we proposed a procedure for constructing the narrow band Hilbert space at a rational magnetic flux ratio $\phi/\phi_0=p/q$ by projecting the zero-field hybrid Wannier basis states (hWS) onto eigenstates of the magnetic translation group (MTG). Such hWSs are Bloch extended along one direction and exponentially localized along the other, and can always be constructed without topological obstruction \cite{Resta2001,Rui2011,Mazari2012}. The pair of hWSs within a valley of TBG carry $\pm 1$ Chern numbers, which are manifested in the intra-moir\'e-unit-cell shift of the averaged position along the localization direction when the Bloch wavenumber along the extended direction is changed \cite{Kang2020a}. We demonstrated that the wavefunctions generated from this projection procedure have a good overlap with the exact wavefunctions obtained using the LL approach at low $\bB$, while being a much more efficient numerical procedure than the LL approach for addressing interaction effects \cite{XW2022}.   

In this work, we present a detailed discussion of the procedure developed in Ref.~\cite{XW2022} and apply it to revisit the non-interacting Hofstadter spectra for square and triangular lattice potentials, where the lowest energy Bloch bands at zero field are topologically trivial. We make quantitative comparisons to the exact LL approach to demonstrate the projection method's regime of validity, as well as derive the Peierls substitution and Harper equation studied extensively in the literature. The paper is organized as follows: In section II we briefly discuss the LL approach and MTG eigenstates. In section III we elaborate on the hWS and how to construct complete and orthonormal set of MTG eigenstates for a subset of Bloch bands at a finite magnetic field. We derive the Peierls substitution as a limiting case when the magnetic length [defined below Eq.~(\ref{eq:ll_wavefunction})] is much longer than the spatial support of the hWS, and we rederive the Harper equation directly using the hWS approach. The summary is provided in section IV.

\section{Landau level approach}
We begin with a brief review of the Landau level (LL) approach for addressing the magnetic field effects on electrons moving in a 2D periodic lattice potential. For notational convenience, we introduce the two lattice vectors as $\ba_1$ and $\ba_2$, and two reciprocal lattice vectors as $\bg_1$ and $\bg_2$. They satisfy the relation $\ba_i\cdot\bg_j=2\pi \delta_{ij}$. For a generic 2D lattice $\ba_1$ and $\ba_2$ are not required to be orthogonal [see e.g. Fig.~\ref{fig:hws}(d)]. We further work with the Landau gauge $\bA=Bx\mathbf{e}_y$ where $B$ is the magnetic field in the out-of-plane $\mathbf{e}_z$-direction, and $\mathbf{e}_y$ ($\mathbf{e}_x$) is along the direction parallel (perpendicular) to $\ba_2$; $\mathbf{e}_{i=x,y,z}$ are unit vectors. The LL wavefunctions are given by
\begin{equation} \label{eq:ll_wavefunction}
\ket{\psi_{n}(k_y)} = \frac{1}{\sqrt{N_2a_2}}e^{ik_yy} \hat{T}(-k_y\ell^2\mathbf{e}_x)\ket{n},
\end{equation}
where $N_2a_2$ is the length of the system along the $\mathbf{e}_y$ direction, $\ell \equiv \sqrt{\frac{\hbar c}{eB}}$ is the magnetic length, $k_y\in\mathbb{R}$ is the momentum quantum number such that $\hat{p}_y\ket{\psi_{n}(k_y)}=\hbar k_y\ket{\psi_{n}(k_y)}$, and $\ket{n}$ is the $n$-th eigenstate for a 1D harmonic oscillator,
\begin{equation}
    \bra{\br}\ket{n} = \frac{1}{\pi^{1/4}\sqrt{2^nn!}}e^{-\frac{x^2}{2\ell^2}}H_n\left(\frac{x}{\ell}\right),
\end{equation} 
where $H_n(x)$ is the Hermite polynomial. The operator $\hat{T}(\br_0)= e^{-i \br_0\cdot\hat{\bp}/\hbar}$ generates a translation by $-\br_0$, i.e. for a general function $f(\br)$ we have
\begin{equation}
\hat{T}(\br_0)f(\br)=f(\br-\br_0).
\end{equation} 

The LL degeneracy argument proceeds as usual: consider a system of area $N_1\ba_1 \times N_2\ba_2$, such that it extends along $\ba_{i=1,2}$ by $N_{i=1,2}\in \mathbb{Z}$ unit cells. For open boundary conditions along $\ba_1$, the quantum number $k_y$ must satisfy $(k_y\ell^2)_\text{max}-(k_y\ell^2)_\text{min}=N_1a_{1x}$. For periodic boundary condition along $\ba_2$, the separation between adjacent wavevectors is $\delta k_y = \frac{2\pi}{N_2a_2}$. The total LL degeneracy is then $\mathcal{N}=\left[(k_y)_\text{max}-(k_y)_\text{min}\right]/\delta k_y=(N_1N_2)\frac{|\ba_1\times\ba_2|}{2\pi \ell^2}$, or equivalently, the LL degeneracy per unit cell is given as 
\begin{equation} \label{eq:LL_dengeneracy}
\frac{\mathcal{N}}{N_1N_2} = \frac{\phi}{\phi_0},
\end{equation}
where $\phi_0=hc/e$ is the magnetic flux quantum, and $\phi = B|\ba_1\times\ba_2|$ is the magnetic flux through a unit cell. 

\subsection{Eigenstates of magnetic translation group}
The single-electron Hamiltonian $\hat{H}(\br,\hat{\bp}+\frac{e}{c}\bA)$ is invariant under discrete magnetic translations $\comm{\hat{H}}{\hat{t}(\ba_{i})}=0$, where $\hat{t}(\ba_{i=1,2})$ are discrete magnetic translation operators along the $\ba_{i=1,2}$ directions. In the Landau gauge, they are given as:
\begin{align}
\hat{t}({\ba_1}) &= e^{-i\bq_\phi \cdot \br}\hat{T}(\ba_1),\\
\hat{t}(\ba_2) &= \hat{T}(\ba_2),
\end{align}
where we defined the wavevector associated with magnetic scattering 
\begin{equation}
\bq_\phi = \frac{2\pi}{a_2}\frac{\phi}{\phi_0}\mathbf{e}_y = \frac{a_{1x}}{\ell^2}\mathbf{e}_y.
\end{equation}
These operators satisfy: 
\begin{equation}
\hat{t}({\ba_2})\hat{t}({\ba_1}) = e^{i2\pi \frac{\phi}{\phi_0}}\hat{t}(\ba_1)\hat{t}(\ba_2).
\end{equation}
Therefore, if ${\phi}/{\phi_0}= {p}/{q}$ where $p$ and $q$ are coprime integers, \begin{equation}
    \comm{\hat{t}(\ba_1)}{\hat{t}^q(\ba_2)}=0.
\end{equation}
As a result, eigenstates of $\hat{H}(\br,\hat{\bp}+\frac{e}{c}\bA)$ can be chosen to be simultaneous eigenstates of $\hat{t}(\ba_1)$ and $\hat{t}^q(\ba_2)$. 

It is straightforward to show that a complete and orthonormal set of MTG basis states can be constructed from the LL wavefunctions as:
\begin{equation} \label{eq:mtg_ll}
\ket{\Psi_{n,r}(\bk)} = \frac{1}{\sqrt{N}}\sum_{s=-\infty}^{\infty} e^{i2\pi k_1 s} \hat{t}^s(\ba_1)\ket{\psi_{n}\left( \frac{2\pi}{a_2}(k_2+\frac{r}{q})\right)},
\end{equation}
where ${N}$ is a normalization factor. The wavevector $\bk=k_1\bg_1+k_2\bg_2$ resides in the magnetic Brillouin zone defined as $k_1\in[0,1)$ and $k_2\in[0,\frac{1}{q})$, and the integer $r$ labels the magnetic strip $[\frac{r-1}{q},\frac{r}{q})$ along the $\bg_2$ direction. The independent basis states are defined for $r=0,\dots,p-1$, because (see appendix A):
\begin{equation}
    \ket{\Psi_{n,r+p}(\bk)}  = e^{i2\pi\left( k_1-(k_2+\frac{r+p}{q})\frac{a_{1y}}{a_2} \right)} \ket{\Psi_{n,r}(\bk)}.
\end{equation}

It is straightforward to check that the states in Eq.~(\ref{eq:mtg_ll}) are MTG eigenstates, i.e.,
\begin{align}
    \hat{t}(\ba_1)\ket{\Psi_{n,r}(\bk)} &= e^{-i2\pi k_1}\ket{\Psi_{n,r}(\bk)},\\
    \hat{t}^q(\ba_2)\ket{\Psi_{n,r}(\bk)} &= e^{-i2\pi q k_2}\ket{\Psi_{n,r}(\bk)}.
\end{align}
They also satisfy the orthonormality condition: 
\begin{equation}
   \bra{\Psi_{n_1,r_1}(\bk)} \ket{\Psi_{n_2,r_2}(\bp)} = \delta_{n_1,n_2}\delta_{r_1,r_2}\delta_{\bk,\bp}.
\end{equation}

In the absence of periodic lattice potential, the LL degeneracy per unit cell is given by $p\cdot 1 \cdot \frac{1}{q}$ consistent with Eq.~(\ref{eq:LL_dengeneracy}). Here $p$ comes from the degeneracy of quantum number $r$, and a fully occupied magnetic Brillouin zone corresponds to $1\cdot\frac{1}{q}$ of the zero-field Brillouin zone occupied, i.e. the fraction of one particle per unit cell. 

Note that $\hat{t}(\ba_2)$ acts non-trivially on the LL-based MTG eigenstates, and because
\begin{equation}
\hat{t}(\ba_2) \ket{\Psi_{n,r}(\bk)} = e^{-i2\pi \left( k_2+\frac{r}{q}\right)}\ket{\Psi_{n,r}(\bk+\frac{p}{q}\bg_1)},
\end{equation}
$\hat{t}(\ba_2)\ket{\Psi_{n,r}(\bk)}$ is an MTG eigenstate at wavevector $\bk+\frac{p}{q}\bg_1$. 

Without loss of clarity from now on we use $\ket{\psi_n(k_2+\frac{r}{q})}$ for convenience, to denote a LL wavefunction with wavenumber $\frac{2\pi}{a_2}\left(k_2+\frac{r}{q}\right)$.

\subsection{Matrix elements of the Hamiltonian}
We study the matrix elements of the continuum single electron Hamiltonian given by: 
\begin{equation}
    \hat{H}(\br,\hat{\bp}+\frac{e}{c}\bA) = \frac{(\hat{\bp}+\frac{e}{c}\bA)^2}{2m_e}+V(\br),
\end{equation}
where $e>0$ is the electric charge, and $V(\br)=\sum_{\bg} V_\bg e^{i\bg\cdot\br}$ is the periodic lattice potential, where $\bg=m\bg_1+n\bg_2$ are reciprocal lattice vectors, $m,n\in \mathbb{Z}$.

The matrix elements of the kinetic energy in the LL-based MTG eigenstate basis [Eq.~(\ref{eq:mtg_ll})] can be straightforwardly calculated by rewriting $\hat{\pi}_x = \frac{\hbar}{\sqrt{2}\ell}(a+a^\dagger)$ and $\hat{\pi}_y = \frac{\hbar}{i\sqrt{2}\ell}(a^\dagger-a)$, where $\hat{\vec{\pi}}=\hat{\bp}+\frac{e}{c}\bA$ is the canonical momentum, and $a$ is the harmonic oscillator lowering operator.

The matrix elements of a general operator of the form $\hat{O}_\bq = \mathcal{O}_\bq e^{i\bq\cdot\br}$ can be calculated as follows \cite{XW2022}: 
\begin{equation}
\begin{split}
 &(\hat{O}_\bq)_{nr_1,mr_2}(\bk,\bp)  \equiv \bra{\Psi_{n,r_1}(\bk)} \hat{O}_\bq \ket{\Psi_{m,r_2}(\bp)} \\
    = & \mathcal{O}_\bq \delta_{p_1,[k_1-q_1]_1}\sum_{s=-\infty}^{\infty}\delta_{\tilde{p}_y-sq_\phi,\tilde{k}_y-q_y}\\
    \times & e^{i2\pi p_1 s}e^{-is\tilde{p}_ya_{1y}}e^{i\frac{s(s-1)}{2}\bq_\phi\cdot\ba_1}e^{-iq_x\tilde{k}_y\ell^2}e^{\frac{i}{2}q_xq_y\ell^2}\\
    \times & \bra{n}e^{c_-a+c_+a^\dagger}\ket{m},
\end{split}
\end{equation}
where we have defined $c_\pm = i\frac{\ell}{\sqrt{2}}(q_x\mp iq_y)$, and:
\begin{equation}
    \tilde{\bk} = \bk+\frac{r_1}{q}\bg_2, \ \tilde{\bp} = \bp+\frac{r_2}{q}\bg_2.
\end{equation}
$\tilde{k}_y$ and $\tilde{p}_y$ are defined as $\tilde{\bk}\cdot\mathbf{e}_y$ and $\tilde{\bp}\cdot\mathbf{e}_y$, respectively. 
The notation $[b]_a$ represents $b$ modulo $a$, with $a>0$. The expression in the last line is calculated as: 
\begin{equation}
\begin{split}
    & \bra{n}e^{c_-a+c_+a^\dagger}\ket{m} \\ 
    = &
\begin{cases}
    e^{\frac{1}{2}c_+c_-}\sqrt{\frac{m!}{n!}}(c_+)^{n-m}L_m^{n-m}(-c_+c_-) &  \text{ for } n\ge m,\\
    e^{\frac{1}{2}c_+c_-}\sqrt{\frac{n!}{m!}}(c_-)^{m-n}L_n^{m-n}(-c_+c_-) &  \text{ for } n< m,
\end{cases}
\end{split}
\end{equation}
where $$L_{n}^{k}(x)= \sum_{m=0}^{n}(-x)^m\frac{(n+k)!}{(n-m)!(k+m)!m!}$$ is the associated Laguerre polynomial. Note that any operator of the form $e^{i\bq\cdot\br}$ is a dense matrix in the LL indices $\{m,n\}$. This poses numerical challenges at low magnetic flux ratios when the upper LL cutoff is large. 

The eigenstates and eigenenergies of the single electron Hamiltonian in a magnetic field and periodic lattice potential can now be solved by diagonalizing the matrix Hamiltonian in the LL-based MTG basis. We briefly discuss the degeneracy of energy levels. Consider an energy eigenstate at a momentum $\bk$ inside the magnetic Brillouin zone, such that: 
\begin{equation} \label{eq:mtg_eigenstates}
    \hat{H}\ket{\tilde{\Psi}_{n}(\bk)} = \varepsilon_{n,\bk}\ket{\tilde{\Psi}_n(\bk)}.
\end{equation}
Making use of the magnetic translation operator $\hat{t}(\ba_2)$, observe that
\begin{equation} 
\begin{split}
    \hat{H}\left( \hat{t}(\ba_2)\ket{\tilde{\Psi}_{n}(\bk)} \right)& = \hat{t}(\ba_2)\hat{H}\ket{\tilde{\Psi}_{n}(\bk)} \\
     &  =\varepsilon_{n,\bk}\left( \hat{t}(\ba_2)\ket{\tilde{\Psi}_{n}(\bk)} \right).
\end{split}
\end{equation}
The state $\hat{t}(\ba_2)\ket{\tilde{\Psi}_{n}(\bk)}$ is therefore also an energy eigenstate at $\varepsilon_{n,\bk}$, but at the wavevector $\bk+\frac{p}{q}\bg_1$. As a result each energy level is at least $q$-fold degenerate, and therefore the dispersion of the resulting magnetic subbands is effectively restricted to the wavevector domain $[0,\frac{1}{q})\times[0,\frac{1}{q})$. 

The LL approach is an exact method, limited in practice only by the truncation of the upper LL index. In the free electron case, retaining $N_c$ LLs allows an accurate representation of the energetics up to $W\approx  \frac{1}{\lambda} N_c\frac{\hbar^2}{m_e\ell^2}$, where $\lambda>1$
is a parameter dependent on the strength of the lattice potential as discussed in the introduction. In the low field limit, the LLs become dense, and a larger LL index is therefore necessary to describe the magnetic subbands emanating from the $\bB=0$ Bloch bands up to the energy $W$. This makes the LL approach both inefficient and not intuitive to study the low field physics, and an alternative approach that bridges the zero field and finite field Hilbert space is preferable. As we show in the next section, this is achieved by projecting the hybrid Wannier basis states (hWSs) -- which form the basis of the $\bB=0$ Hilbert space -- onto representations of the MTG.

\section{Hybrid Wannier approach}
\subsection{Hybrid Wannier states at $\bB=0$}
\begin{figure*}
\includegraphics[width=\textwidth]{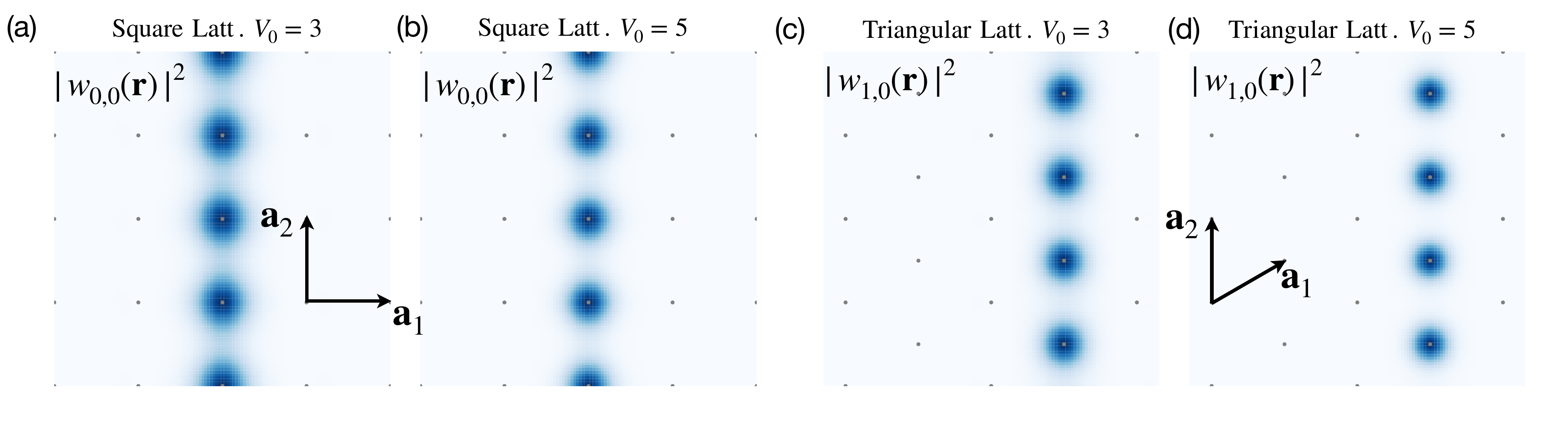}
\caption{\label{fig:hws} Hybrid Wannier states $w_{n_0,k_2}(\br)$ of the lowest energy Bloch band for a square lattice (a,b) and triangular lattice (c,d) potential. In both cases we used the potential $V(\br)=-V_0\sum_{\bg}e^{i\bg\cdot\br}$. We set $\hbar^2/m_ea^2=1$, where $m_e$ is the bare electron mass, and $a$ the lattice constant. For square lattice, $\bg=\pm \bg_1,\pm \bg_2$, whereas for triangular lattice, $\bg=\pm \bg_1,\pm \bg_2,\pm (\bg_1+\bg_2)$. The states are well localized on the lattice sites, and their spatial support narrows with increasing lattice potential strength.}
\end{figure*}

In the absence of the magnetic field, the energy eigenstates are given by the Bloch states 
\begin{equation}
    \psi_{n,\bk}(\br)=\frac{1}{\sqrt{\mathcal{A}}}e^{i\bk\cdot\br}u_{n,\bk}(\br),
\end{equation}
where $\mathcal{A}$ is the area of the 2D system, $n$ is the band label, $\bk=k_1\bg_1+k_2\bg_2$ is the crystal momentum in the first Brillouin zone [for convenience we choose it to be $k_1,k_2\in[0,1)\times[0,1)$], and $u_{n,\bk}(\br)$ is periodic under discrete lattice translations $\hat{T}(\ba_{i=1,2})$.  The notation $\psi_{n,\bk}(\br)$ should not be confused with the LL wavefunctions discussed in the previous section. One can construct spatially localized basis states by performing unitary transformations on the (extended) Bloch energy eigenstates. For a subset of Bloch bands of interest, we can construct hWSs from the Bloch states even if these bands have non-trivial topology, provided that the gap to the adjacent bands does not close. hWSs are exponentially localized in one direction (say $\ba_1$), and Bloch extended along the other (say $\ba_2$). 
In Fig.~\ref{fig:hws} we show several examples of hWSs for the lowest energy Bloch band on a square and triangular lattice. These states can be constructed \cite{Rui2011,Kang2020a} by diagonalizing the periodic version of the position operator $\exp(i\delta \bk \cdot \br)$ projected onto the Bloch basis from the desired energy bands, where $\delta \bk= \frac{1}{N_1}\bg_1$  hybridizes Bloch states at $\bk$ and $\bk+\delta\bk$. The resulting hWS that is exponentially localized near a column at $n_0\ba_1$ can be expressed as:
\begin{equation} \label{eq:hw_bloch}
    \ket{w_\alpha(n_0,k_2)} = \frac{1}{\sqrt{N_1}}\sum_{n\in {\rm subset}} \sum_{k_1} e^{-i2\pi k_1 n_0} U_{n,\alpha}(\bk)\ket{\psi_{n,\bk}},
\end{equation}
where the summation over $n$ is over a subset of Bloch bands of interest, and $U(\bk)$ is a unitary matrix at every $\bk$.
Under discrete translations the hWSs satisfy 
\begin{align}
    \hat{T}(\ba_1)\ket{w_\alpha(n_0,k_2)} &=\ket{w_\alpha(n_0+1,k_2)}, \\
    \hat{T}(\ba_2)\ket{w_\alpha(n_0,k_2)} &=e^{-i2\pi k_2}\ket{w_\alpha(n_0,k_2)}. 
\end{align}

For topologically non-trivial bands, the hWSs contain information about non-zero Chern number of a Bloch band, which is manifested in the non-trivial evolution of the averaged position $\langle \br\cdot\bg_1\rangle/|\bg_1|$ within the hWSs when $k_2$ is continuously increased from 0 to 1. One example of such hWSs is the pair of narrow bands in TBG for a given valley and spin. For more details we refer interested readers to Refs.~\cite{Kang2020a,XW2022}. 

\subsection{MTG eigenstates from hybrid Wannier states}
The set of hWSs for all Bloch bands forms a complete basis even in finite magnetic field. However it is not useful if we are only interested in the Hofstadter physics of a subset of Bloch bands. 

It is tempting, but wrong, to take the subset of the $\bB=0$ Bloch bands as a basis of the corresponding subset of the $\bB\neq 0$ states. Note that the correct projector at $\bB\neq0$, $\hat{P}_B=\sum_{n}'\sum_{\bk}\ket{\tilde{\Psi}_{n}(\bk)}\bra{\tilde{\Psi}_{n}(\bk)}$ [see Eq.~(\ref{eq:mtg_eigenstates})], where $n$ is summed over the subset of magnetic subbands of interest, is invariant under any integer multiple of magnetic translations, i.e., $\comm{\hat{P}_B}{\hat{t}^{s_1}(\ba_1)\hat{t}^{s_2}(\ba_2)}=0$, where $s_1,s_2\in\mathbb{Z}$. 
However, the $\bB=0$ projector, $\hat{P}=\sum_{n\in \mathrm{subset}}\sum_{\bk}\ket{\psi_{n,\bk}}\bra{\psi_{n,\bk}}$, is not invariant under $\hat{t}^{s_1}(\ba_1)$, because
\begin{eqnarray}\label{eq:projector}
&&\hat{t}^{s_1}(\ba_1)\hat{P}\hat{t}^{-s_1}(\ba_1)=\\
    &&\sum_{mm'\in \mathrm{fullset}} \sum_{\bk} \ket{\psi_{m,\bk}}\bra{\psi_{m',\bk}}
\sum_{n\in \mathrm{subset}}U_{mn}(\bk)U_{nm'}^\dagger(\bk).\nonumber
\end{eqnarray}
Here we defined 
\begin{equation}
    \hat{t}^{s_1}(\ba_1)\ket{\psi_{n,\bk}} = \sum_{m\in {\rm fullset}} U_{mn}(\bk) \ket{\psi_{m,\bk}}.
\end{equation}
Due to the restriction on $n$, the right-hand-side of Eq.~(\ref{eq:projector}) is not equal to $\hat{P}$, the narrow-band projector at $\bB=0$. In other words, the $y$-dependent phase in $\hat{t}(\ba_1)$ takes the states outside of the subset of the zero-field bands of interest. This problem is severe even at low $\bB$ for sufficiently large $s_1$. Therefore $\hat{P}$ is {\it not} a projector onto the states of interest in finite magnetic fields.

In order to construct the correct finite-field Hilbert space, we first construct MTG eigenstates from the hWSs
\begin{equation} \label{eq:mtg_hw}
    \ket{W_{\alpha,r}(\bk)} = \frac{1}{\sqrt{N_1}}\sum_{s=-\frac{N_1}{2}}^{\frac{N_1}{2}}e^{i2\pi k_1 s} \hat{t}^{s}({\ba_1})\ket{w_\alpha(0,k_2+\frac{r}{q})}.
\end{equation}
Unlike in the previous subsection, here $k_2\in[0,\frac{1}{q})$, and 
$r=0,\dots,q-1$. The choice of hWSs  at $n_0=0$ is motivated by the fact that the vector potential in the Landau gauge vanishes at the origin. Therefore, $\ket{w_\alpha(0,k_2+\frac{r}{q})}$ must have a large overlap onto the subbset of the $\bB\neq 0$ Hilbert space with the similar energy at small $B$, i.e. onto the magnetic subbands emanating from the $\bB=0$ bands of interest. The rest of the basis can be conveniently obtained by using the magnetic translation operator which moves hWSs along the $\ba_1$ direction, while also attaching a phase due to the vector potential. 

It is straightfoward to check that $\ket{W_{\alpha,r}(\bk)}$ are indeed MTG eigenstates, i.e., 
\begin{align}
\hat{t}({\ba_1})\ket{W_{\alpha,r}(\bk)} & = e^{-i2\pi k_1}\ket{W_{\alpha,r}(\bk)},\\
\hat{t}^{q}({\ba_2})\ket{W_{\alpha,r}(\bk)} & = e^{-i2\pi q k_2}\ket{W_{\alpha,r}(\bk)}.
\end{align}

\begin{figure*}[t]
\includegraphics[width=\linewidth]{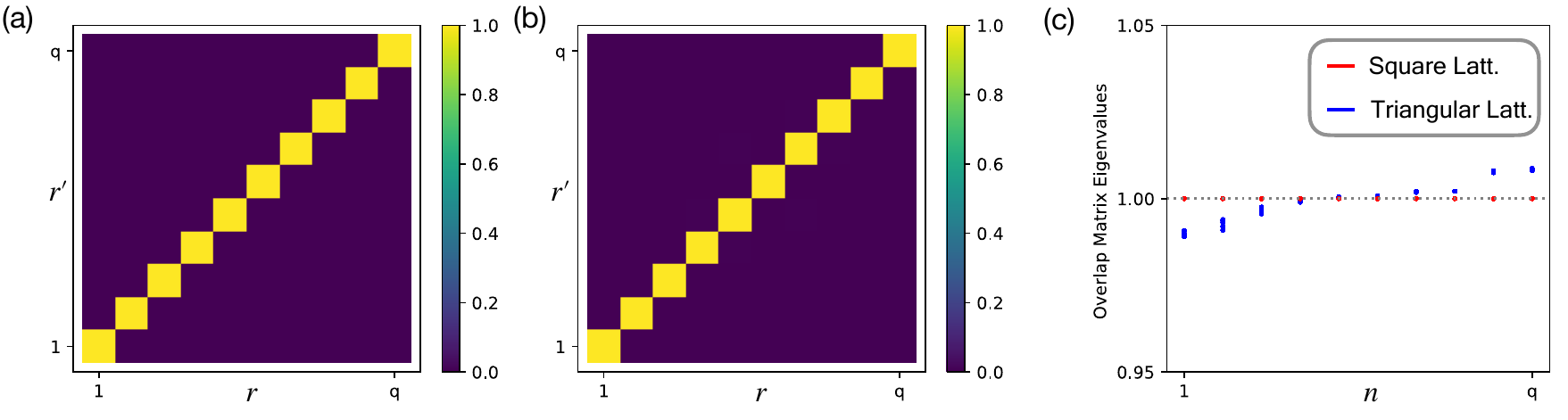}
\caption{\label{fig:overlap_eigvals} Typical absolute value of the matrix elements of the overlap matrix [Eq.~(\ref{eq:hWS_overlap})], calculated at flux $\phi/\phi_0=1/10$ for (a) square and (b) triangular lattice potentials respectively with $V_0=3$. We set the energy scale $\hbar^2/m_ea^2=1$. (c) Eigenvalues of the overlap matrix. The axes in (a,b) correspond to the basis index $r,r'$, and in (c) the ordered eigenvalue index. If the MTG basis states generated from hWSs constitute a complete and orthonormal basis, the overlap matrix is an identity matrix and all eigenvalues are equal to 1. While this is true for square lattice potential, there are small deviations from orthonormality for the case of triangular lattice potential. As a result a further orthonormalization procedure outlined in the text is needed.}
\end{figure*}

To address the completeness and orthonormality of the wavefunctions defined in Eq.~(\ref{eq:mtg_hw}), we define an overlap matrix 
\begin{equation}
    \Lambda_{\alpha r_1,\beta r_2}(\bk)\equiv \bra{W_{\alpha,r_1}(\bk)}\ket{W_{\beta,r_2}(\bk)}.
\end{equation}
Note that states with different $\bk$ are automatically orthogonal due to different eigenvalues under $\hat{t}(\ba_{i=1,2})$. If these wavefunctions represent a complete and orthonormal set, the overlap matrix should be an identity matrix. 
The overlap matrix can be calculated as follows:
\begin{widetext}
\begin{equation} \label{eq:hWS_overlap}
    \begin{split}
        \Lambda_{\alpha r_1,\beta r_2}(\bk) =  & \sum_{s} e^{i2\pi k_1 s}\bra{w_\alpha(0,k_2+\frac{r_1}{q})} \hat{t}^s({\ba_1})\ket{w_\beta(0,k_2+\frac{r_2}{q})}\\
        = & \sum_{s} e^{i2\pi k_1 s} e^{i\frac{s(s-1)}{2}\bq_\phi\cdot\ba_1}\bra{w_\alpha(0,\tilde{k}_2)} e^{-is\bq_\phi\cdot\br}\ket{w_\beta(s,\tilde{p}_2)},
    \end{split}
\end{equation}
\end{widetext}
where on second line we defined $\tilde{k}_2=k_2+\frac{r_1}{q}$ and $\tilde{p}_2=k_2+\frac{r_2}{q}$ ($\tilde{k}_2,\tilde{p}_2\in[0,1)$), and used the operator identity in Eq.~(\ref{eq:magnetic_translation_expansion}). 

If the hWSs represent topologically trivial Bloch bands (e.g., Fig.~\ref{fig:hws}), the overlap of two hWSs is exponentially suppressed unless they have the same localization center, i.e. unless $s=0$ in Eq.~(\ref{eq:hWS_overlap}). Therefore $\bra{W_{\alpha,r_1}(\bk)}\ket{W_{\beta,r_2}(\bp)}\approx \delta_{\alpha,\beta}\delta_{\bk,\bp}\delta_{r_1,r_2}$ with  exponential accuracy. This is illustrated in the Fig.~\ref{fig:overlap_eigvals} for the square and triangular lattice potentials. Even if $\Lambda(\bk)$ is not an identity matrix, a complete and orthonormal basis set can be generated by eigen-decomposition: $U^\dagger \Lambda U= D$, and redefining a new set of basis states as: 
\begin{equation} \label{eq:mtg_hw_norm}
    \ket{V_a(\bk)} = \sum_{\alpha,r} \ket{W_{\alpha,r}(\bk)} \left( U \frac{1}{\sqrt{D}}\right)_{\alpha r,a}. 
\end{equation}

If hWSs represent topological bands with finite Chern numbers, hWSs with different localization centers have a finite spatial overlap, and the overlap matrix in Eq.~(\ref{eq:hWS_overlap}) strongly deviates from an identity matrix. Importantly, as discussed in Ref.~\cite{XW2022}, for hWSs with Chern numbers $\pm 1$, the number of independent MTG eigenstates should increase/decrease according to the Streda formula. This manifests in the rank deficiency of the overlap matrix. A procedure that makes use of a different choice of the localization center $n_0$ (still close to $0$ where the $\bA$ is small) was proposed in Ref.~\cite{XW2022}, where it was shown to resolve this issue. 

\begin{figure}
\includegraphics[width=\linewidth]{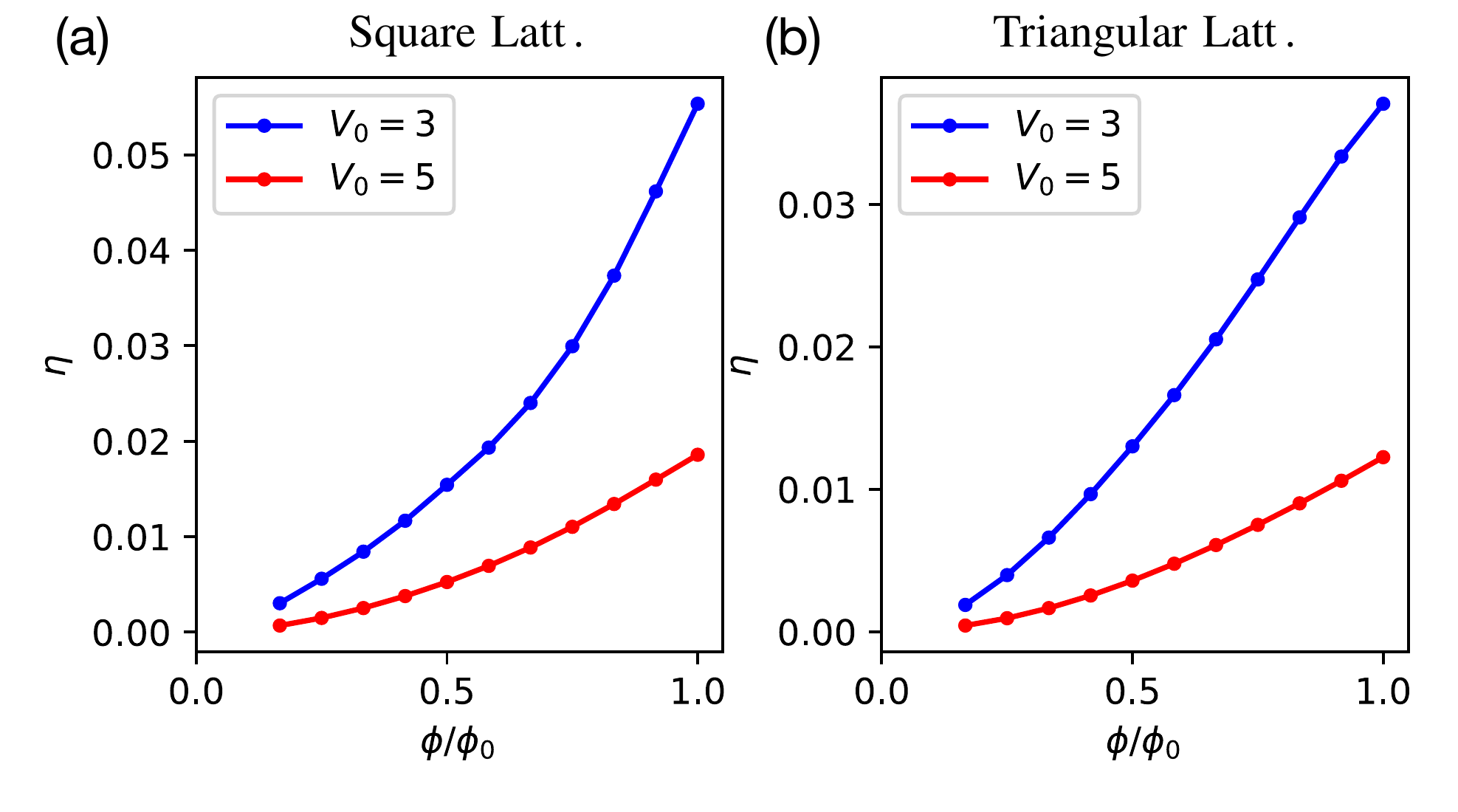}
\caption{\label{fig:spillover} The amount of spillover $\eta$ of MTG basis states into the Hilbert space of remote Hofstadter bands for square and triangular lattice potential. We set the energy scale $\hbar^2/m_ea^2=1$. The spillover decreases with increasing lattice potential strength, and increases with increasing magnetic flux. }
\end{figure}

The next question is how well do these basis states describe finite field Hilbert space of interest. 
To quantify the amount of spillover into remote magnetic subbands, we can expand Eq.~(\ref{eq:mtg_hw_norm}) in the exact wavefunctions obtained based on the LL approach,
\begin{equation} \label{eq:hws_ll}
\begin{split}
    \ket{V_{a}(\bk)} & = \sum_{n\in \mathrm{active}} M_{n,a}(\bk)\ket{\tilde{\Psi}_{n}(\bk)} \\
    & + \sum_{n'\in \mathrm{remote}} M_{n',a}(\bk)\ket{\tilde{\Psi}_{n'}(\bk)},
\end{split}
\end{equation}
where $\ket{\tilde{\Psi}_{n}(\bk)}$ are the exact eigenstates obtained from the LL approach, and $M_{n,a}(\bk)$ are the expansion coefficients. Define:
\begin{equation}
    \eta_a \equiv \frac{1}{N_1}\frac{1}{N_2/q}\sum_{\bk}\sum_{n'\in \mathrm{remote}}|M_{n',a}(\bk)|^2.
\end{equation}
 Then $\eta_a\in[0,1]$ characterizes the amount of spillover into remote magnetic subbands. In Fig.~\ref{fig:spillover} we plot $\eta=\eta_{a=1}$ versus $\phi/\phi_0$ for the lowest energy Bloch band for square and triangular lattice potentials. The amount of spillover decreases with increasing strength of the lattice potential, and remains small even for reasonably large magnetic flux.

\subsection{Peierls factor and MTG basis states}
To gain a better insight into the MTG basis states obtained from zero field hWSs, let us consider an example of an isolated and topologically trivial band, where 2D Wannierization can be achieved. The hWSs are related to 2D Wannier orbitals via Fourier transform along the $\ba_2$ axis, i.e.,
\begin{equation} \label{eq:hws_2dwannier}
    \bra{\br}\ket{w(n_0,k_2)} = \frac{1}{\sqrt{N_2}}\sum_{\bR}\phi_\bR(\br)\delta_{\bR\cdot\bg_1,2\pi n_0} e^{ik_2\bg_2\cdot\bR}.
\end{equation} 
$\phi_\bR(\br)$ denotes a 2D localized Wannier orbital at site $\bR=m\ba_1+n\ba_2$, with the normalization condition $\int \mathrm{d}^2\br \phi_\bR^*(\br)\phi_{\bR'}(\br)=\delta_{\bR,\bR'}$. The MTG basis states can then be written as
\begin{equation} \label{eq:mtg_2dwannier}
\begin{split}
    \bra{\br}\ket{W_r(\bk)} =  \frac{1}{\sqrt{N_1N_2}}\sum_{\bR} e^{i\tilde{\bk}\cdot\bR} \hat{t}^{m}({\ba_1})\hat{t}^n({\ba_2}) \phi_\mathbf{0}(\br),
\end{split}
\end{equation}
where we have defined $\tilde{\bk} = k_1\bg_1 +( k_2 + \frac{r}{q})\bg_2$. Compared to the zero-field Bloch states, the MTG states are obtained by acting on the 2D Wannier orbital at site $\bR=\mathbf{0}$ with non-commuting magnetic translation operators instead of the usual lattice translation operators. 

We  make a connection to the Peierls factor used in the literature \cite{Luttinger1951}, where the phase $\exp\left(-i\frac{e}{\hbar c}\int_{\bR}^\br \mathrm{d}\br' \cdot \bA(\br')\right)$ is attached to the 2D ($\bB=0$) Wannier orbital $\phi_{\bR}(\br)$ with the integration along the straight line from $\bR$ to $\br$. 
Expanding out the magnetic translation operators in Eq.~(\ref{eq:mtg_2dwannier}):
\begin{equation}
    \bra{\br}\ket{W_r(\bk)} =  \frac{1}{\sqrt{N_1N_2}}\sum_{\bR} e^{i\tilde{\bk}\cdot\bR} e^{i\frac{m(m-1)}{2}\bq_\phi\cdot\ba_1}e^{-im\bq_\phi\cdot\br} \phi_\bR(\br).
\end{equation}
If the Wannier orbitals are exponentially localized near the lattice sites, and the magnetic length is much longer than their spatial extent, then we can take $|\br-\bR|/\ell \ll 1$ and  approximate
\begin{equation}
    e^{-i m\bq_\phi \cdot \br}\phi_\bR(\br) \approx e^{-i m\bq_\phi \cdot \bR} \left [ e^{-i\frac{e}{\hbar c}\int_\bR^{\br}\mathrm{d}\br'\cdot \bA(\br')}\phi_\bR(\br)  \right],
\end{equation}
where we used 
\begin{equation}
\begin{split}
    & i\frac{e}{\hbar c}\int_\bR^{\br}\mathrm{d}\br'\cdot\bA(\br')\\
    = & \frac{i}{\ell^2}R_x(y-R_y) +\frac{i}{2\ell^2}(x-R_x)(y-R_y)\\
    = & i m \frac{a_{1x}}{\ell^2} (y-R_y) + \frac{i}{2\ell^2}(x-R_x)(y-R_y)\\
    = & i m \bq_\phi\cdot(\br-\bR) + O(|\br-\bR|^2/\ell^2).
\end{split}
\end{equation}
This explicitly demonstrates the link between the MTG basis states and the Wannier orbitals with a Peierls factor. 

\subsection{Matrix elements of the Hamiltonian}
Here we work out the matrix elements of the single-electron Hamiltonian in the MTG basis states defined in Eq.~(\ref{eq:mtg_hw}). Matrix elements with respect to the orthonormal basis $\ket{V_a(\bk)}$ can be obtained via the basis transformation in Eq.~(\ref{eq:mtg_hw_norm}).  Specifically:
\begin{equation}\label{eq:Hmat_hws}
    \begin{split}
     & \bra{W_{\alpha,r_1}(\bk)}\hat{H}\ket{W_{\beta,r_2}(\bk)}  = \sum_{s}e^{i2\pi k_1 s} \\
     \times &  e^{i\frac{s(s-1)}{2}\bq_\phi\cdot\ba_1}  \bra{w_{\alpha}(0,\tilde{k}_2)}\hat{H} e^{-is\bq_\phi\cdot\br}\ket{w_{\beta}(s,\tilde{p}_2)},
    \end{split}
\end{equation}
where we defined $\tilde{k}_2=k_2+\frac{r_1}{q}$ and $\tilde{p}_2=k_2+\frac{r_2}{q}$ ($\tilde{k}_2,\tilde{p}_2\in[0,1)$). This expression is obtained by expanding the MTG basis states using Eq.~(\ref{eq:mtg_hw}), and making use of the operator identity (see appendix A): 
\begin{equation}
    \hat{t}^s(\ba_1) = e^{i\frac{s(s-1)}{2}\bq_\phi\cdot\ba_1}e^{-is\bq_\phi\cdot\br} \hat{T}^{s}(\ba_1).
\end{equation}

We hereby split the full single-electron Hamiltonian into the zero-field component $\hat{H}_0$ and the finite field component $\hat{H}_{\bB}=\hat{H}-\hat{H}_0$ respectively, and calculate each term separately. 

\subsubsection{$\hat{H}_0$}
For the zero field Hamiltonian, its matrix elements can be straightforwardly studied as follows: 
\begin{equation} \label{eq:I1}
\begin{split}
         I_1 = & \sum_{s,s_0,\gamma} e^{i2\pi k_1 s} e^{i\frac{s(s-1)}{2}\bq_\phi\cdot\ba_1} \\
         & \times \bra{w_{\alpha}(0,\tilde{k}_2)}\hat{H}_0 \ket{w_{\gamma}(s_0,\tilde{k}_2)}\\
         & \times \bra{w_{\gamma}(s_0,\tilde{k}_2)}e^{-is\bq_\phi\cdot\br}\ket{w_{\beta}(s,\tilde{p}_2)},
        \end{split}
\end{equation}
where we have inserted the narrow band projector $\sum_{s_0 \gamma}\ket{w_{\gamma}(s_0,\tilde{k}_2)}\bra{w_{\gamma}(s_0,\tilde{k}_2)}$. 
The second line can be calculated by writing down hWSs in the Bloch band basis using Eq.~(\ref{eq:hw_bloch}): 
\begin{equation}\label{eq:I1_1}
\begin{split}
    & \bra{w_{\alpha}(0,\tilde{k}_2)}\hat{H}_0 \ket{w_{\gamma}(s_0,\tilde{k}_2)} \\
= & \frac{1}{N_1}\sum_{\tilde{k}_1,n} e^{-i2\pi \tilde{k}_1s_0} \left[ U^\dagger_{\alpha, n}(\tilde{\bk})\varepsilon_{n,\tilde{\bk}}U_{n,\gamma}(\tilde{\bk}) \right],
\end{split}
\end{equation}
where in this section we have redefined $\tilde{\bk}=\tilde{k}_1\bg_1+\tilde{k}_2\bg_2$ ($\tilde{k}_1,\tilde{k}_2\in[0,1)$). It leads to 1D hopping between hWSs at different localization centers with the same wavevector $\tilde{k}_2$. The last line in Eq.~(\ref{eq:I1}) can be calculated as: 
\begin{equation} \label{eq:I1_2}
    \begin{split}
        & \bra{w_{\gamma}(s_0,\tilde{k}_2)}e^{-is\bq_\phi\cdot\br}\ket{w_{\beta}(s,\tilde{p}_2)}  \\
        = & \frac{1}{N_1}\sum_{\tilde{k}_1,\tilde{p}_1} e^{i2\pi \tilde{k}_1s_0}e^{-i2\pi\tilde{p}_1s}\\
        & \times \sum_{n,m} U^{\dagger}_{\gamma,n}(\tilde{\bk})\bra{\psi_{n}(\tilde{\bk})}e^{-is\bq_\phi\cdot\br}\ket{\psi_{m}(\tilde{\bp})}U_{m,\beta}(\tilde{\bp}).
    \end{split}
\end{equation}
It represents scattering between Bloch eigenstates by wavevector $s\bq_\phi$. Therefore, the matrix elements of the zero-field Hamiltonian in the MTG basis represent a hop along the $\ba_1$ direction, followed by a magnetic scattering by wavevector $s\bq_\phi$. Note that if $\ba_1\cdot\ba_2\neq 0$, i.e., non-orthogonal unit cell vectors such as triangular or honeycomb lattice, magnetic scattering hybridizes Bloch states with wavevectors $\tilde{\bk}=\tilde{k}_1\bg_1+\tilde{k}_2\bg_2$ and $\tilde{\bp}=\tilde{p}_1\bg_1+\tilde{p}_2\bg_2$ such that:  
\begin{equation} \label{eq:I1_3}
\tilde{p}_1 =[\tilde{k}_1+\frac{sp}{q}\frac{a_{1y}}{a_2}]_1,\ r_2=[r_1+sp]_q.
\end{equation}

\begin{figure*}
\includegraphics[width=\linewidth]{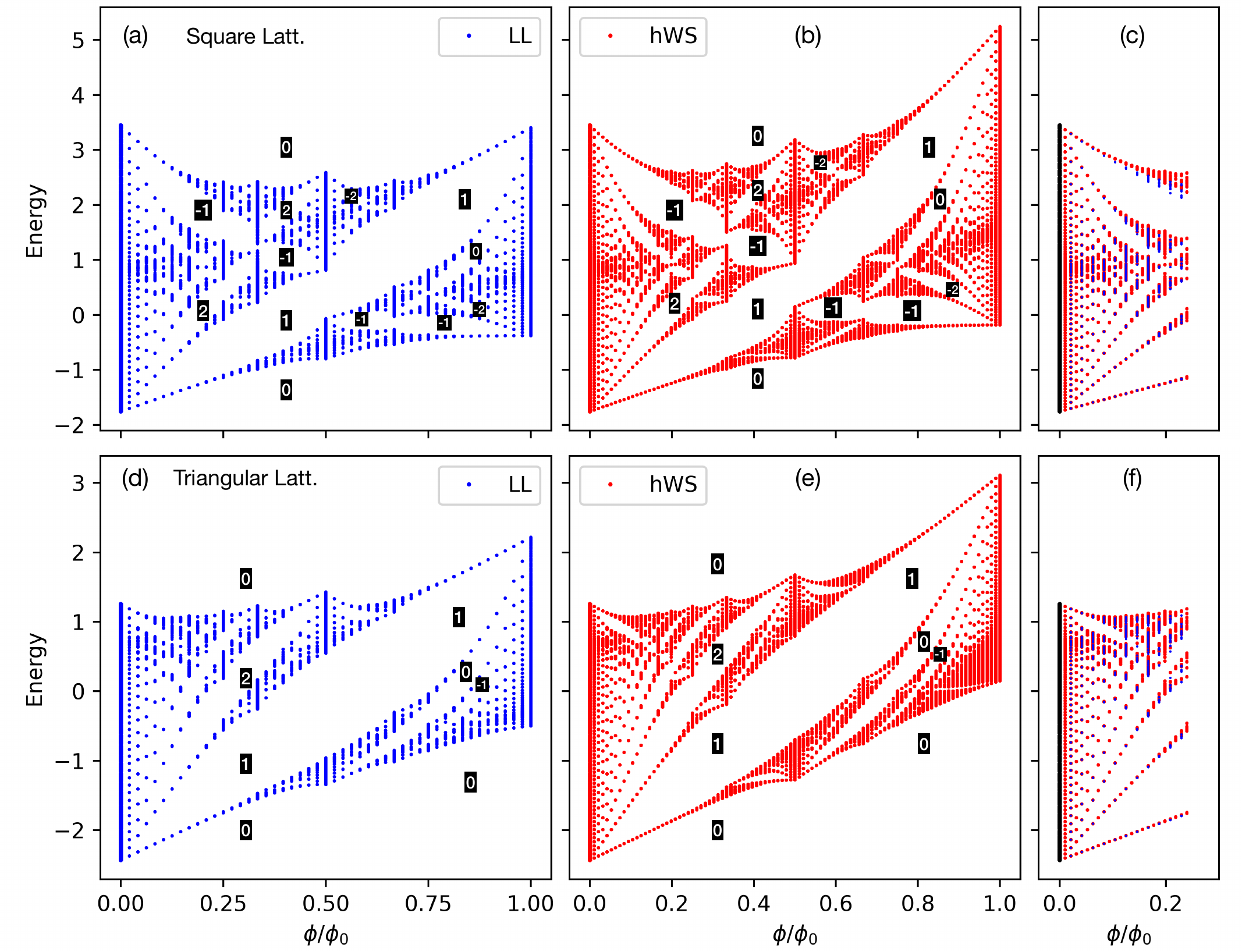}
\caption{\label{fig:hofstadter_comparison}Hofstadter spectra obtained using exact LL approach (blue) and the hWS approach (red). Upper panels are square lattice and lower panels are triangular lattice. A lattice potential strength $V_0=3$ is used. We set the energy scale $\hbar^2/m_ea^2=1$. For LL approach we studied the $p/q$ sequence for $q=48$ and $p=1,\dots48$. For hWS approach we studied the sequence $q=96$ and $p=1,\dots96$. The numbers in figures (a,b,d,e) label the Chern number of the corresponding energy gaps.}
\end{figure*}

\subsubsection{$\hat{H}_{\bB}$} \label{sec:HB}
The finite field term in the Landau gauge is given by $\hat{H}_{\bB}(\br)=\frac{\hbar x\hat{p}_y}{m_e\ell^2}+\frac{\hbar^2x^2}{2m_e\ell^4}$. It contains polynomials of the coordinates, and grows when moving away from the axes origin where $x=0$. Below we work out the matrix elements for the term $\frac{\hbar x\hat{p}_y}{m_e\ell^2}$, and the second term can be calculated in an analogous fashion. Specifically,
\begin{equation}\label{eq:I2}
\begin{split}
    I_2 = & \sum_{s} e^{i2\pi k_1 s} e^{i\frac{s(s-1)}{2}\bq_\phi\cdot\ba_1}\\
     \times & \frac{\hbar^2}{m_e\ell^2} \bra{w_{\alpha}(0,\tilde{k}_2)}\frac{x\hat{p}_y}{\hbar}e^{-is\bq_\phi\cdot\br}\ket{w_{\beta}(s,\tilde{p}_2)}.
\end{split}
\end{equation}

We first note that since the hWSs are exponentially suppressed away from their localization centers, for $s\gg 0$, the term is exponentially small regardless of the operator. Using the estimate $\langle x\rangle \sim a_{1x}$ and $\langle \hat{p}_y\rangle\sim 2\pi \hbar / a_2$, $I_2 \sim \frac{\hbar^2}{m_e\ell^2}$, i.e., cyclotron frequency.  At low fields such that $p/q\ll 1$, the finite field term is smaller than the zero field term by ratio $p/q$. Therefore, unless there are further band flattening effects for the zero-field term (e.g., TBG at magic angle), the finite field term is negligible. 

The second line of Eq.~(\ref{eq:I2}) can be calculated by expanding in the Bloch basis: 
\begin{equation}
\begin{split}
    & \frac{\hbar}{m_e\ell^2 N_1} \sum_{\tilde{k}_1\tilde{p}_1} e^{-i2\pi\tilde{p}_1s}\\
    \times & \sum_{n,m} U^\dagger_{\alpha,n}(\tilde{\bk})U_{m,\beta}(\tilde{\bp}) \bra{\psi_n(\tilde{\bk})}\frac{x\hat{p}_y}{\hbar}e^{-is\bq_\phi\cdot\br}\ket{\psi_m(\tilde{\bp})},
\end{split}
\end{equation}
where second line can be explicitly calculated in the plane wave basis: 
\begin{equation}
    \begin{split}
        &\bra{\psi_n(\tilde{\bk})}\frac{x\hat{p}_y}{\hbar}e^{-is\bq_\phi\cdot\br}\ket{\psi_m(\tilde{\bp})}\\
        = & \sum_{\bg,\bg'} u_{n\bg}^*(\tilde{\bk})u_{m\bg'}(\tilde{\bp}) \frac{1}{\mathcal{A}}\int\mathrm{d}^2\br\\
         & e^{-i(\bg+\tilde{\bk})\cdot\br}\left[x\left(g_y+\tilde{k}_y\right)\right]e^{-is\bq_\phi\cdot\br}e^{i(\bg'+\tilde{\bp})\cdot\br}.
    \end{split}
\end{equation}
Here  we have expanded the periodic part of the Bloch states in a Fourier series: $u_{n\tilde{\bk}}(\br)=\sum_{\bg}e^{i\bg\cdot\br}u_{n\bg}(\tilde{\bk})$, and $\mathcal{A}=N_1N_2a_{1x}a_{2y}$ is the area of the system. It is important that the real space integral $\int \mathrm{d}^2\br[...]$ must be placed in a box to avoid the revivals of the hWSs along the $\mathbf{e}_x$ direction at the boundary of a torus. In practice we choose the integration domain as $x\in[-\frac{N_1a_{1x}}{2},-\frac{N_1a_{1x}}{2})$ (see appendix B for details). 

\subsection{Comparing Hofstadter spectra}
In Fig.~\ref{fig:hofstadter_comparison} we show the comparison of the Hofstadter spectra for the lowest energy Bloch band with square [Figs.~\ref{fig:hofstadter_comparison}(a,b,c)] and triangular [Figs.~\ref{fig:hofstadter_comparison}(d,e,f)] lattice potentials respectively, calculated using the exact LL approach and the approximate hWS approach. For the parameters used in the calculation, the hWSs are well localized within a unit cell (see Fig.~\ref{fig:hws}), and there is a good quantitative agreement of the Hofstadter spectra up to $\phi/\phi_0\approx 0.2$. At higher magnetic flux, the MTG states generated from hWSs have a larger spillover onto remote magnetic subbands, and lead to an overall upward shift of the Hofstadter spectra when compared to LL calculations. Despite the quantitative differences in the Hofstadter spectra, the prominent magnetic subbands and the Chern numbers associated with gaps are correctly captured via the hWS method all the way to  $\phi/\phi_0=1$, as indicated by the labels in Figs.~\ref{fig:hofstadter_comparison}(a,b,d,e).  

\subsection{Harper equation}
Eq.~(\ref{eq:Hmat_hws}) gives a general procedure for calculating the matrix elements of the single-electron Hamiltonian for a subset of Bloch bands isolated from the rest, by making use of the hWSs at zero field and projecting onto representations of the MTG at finite field. Here we elaborate on the case of a square lattice potential and an isolated band with trivial topology. As mentioned, in this case the hWSs are just the 1D Fourier transforms of the Bloch eigenstates with a smooth gauge in the Brillouin zone and the relation to 2D Wannier orbitals is described in Eq.~(\ref{eq:hws_2dwannier}). Below we show that the Harper equation of Hofstadter \cite{Hofstadter1976} is recovered as long as the magnetic length $\ell$ is much longer than the localization length $\xi$ of the Wannier orbitals. We express the left hand side of Eq.~(\ref{eq:I1_2}) using 2D exponentially localized Wannier orbitals,
\begin{widetext}
\begin{equation} \label{eq:HarperI2}
\begin{split}
    & \bra{w(s_0,\tilde{k}_2)}e^{-is\bq_\phi\cdot\br}\ket{w(s,\tilde{p}_2)} \\
= & \frac{1}{N_2}\sum_{n_1,n_2}\left[ \int \mathrm{d}^2\br \phi_{\bR_1}^*(\br) \phi_{\bR_2}(\br)e^{-is\bq_\phi\cdot(\br-\bR_2)}\right] e^{-is\bq_\phi\cdot\bR_2} e^{-i2\pi\tilde{k}_2n_1}e^{i2\pi\tilde{p}_2n_2},
\end{split}
\end{equation}
\end{widetext}
where we defined $\bR_1=s_0\ba_1+n_1\ba_2$ and $\bR_2=s\ba_1+n_2\ba_2$. Since Wannier orbitals are exponentially localized, $s_0$ and $s_1$ have to be close to each other. Making the assumption that the 1D hopping is short ranged, Eq.~(\ref{eq:I1_1}) forces  $s_0$ to be near 0, therefore $s$ is also close to 0. As a result, the factor $e^{-is\bq_\phi\cdot(\br-\bR_2)}$ is slowly varying over the lengthscale $\sim\xi$, the spatial extent of the exponentially localized Wannier orbital, and can be Taylor expanded in the real space integration in the bracket. The integral can therefore be expanded in power series of $\xi/\ell$ as, 
\begin{equation}
    \begin{split}
        & \int \mathrm{d}^2\br \phi_{\bR_1}^*(\br) \phi_{\bR_2}(\br)e^{-is\bq_\phi\cdot(\br-\bR_2)} \\
        \approx & \int \mathrm{d}^2\br \phi_{\bR_1}^*(\br) \phi_{\bR_2}(\br)
        \left[ 1-is\bq_\phi\cdot(\br-\bR_2)+\ldots\right] \\
        = & \delta_{s_0,s}\delta_{n_1,n_2}+O(a_{1x}\xi/\ell^2).
    \end{split}
\end{equation}
 As a result, Eq.~(\ref{eq:HarperI2}) becomes:
\begin{equation}
\begin{split} 
    & \bra{w(s_0,\tilde{k}_2)}e^{-is\bq_\phi\cdot\br}\ket{w(s,\tilde{p}_2)}\\
    = & \delta_{s_0,s}\delta_{[r_1+sp]_q,r_2} + O(a_{1x}\xi/\ell^2),
\end{split}
\end{equation}
 where in the last line we have substituted $\tilde{k}_2=k_2+r_1/q$ and $\tilde{p}_2=k_2+r_2/q$. Keeping leading order term in the above expression, and combining with Eq.~(\ref{eq:I1_1}) we get the following simplification for Eq.~(\ref{eq:I1}): 
\begin{equation} 
\begin{split}
      I_1 \approx & \frac{1}{N_1}\sum_{s}e^{i2\pi k_1s} \sum_{\tilde{k}_1} e^{-i2\pi \tilde{k}_1s}\varepsilon_{\tilde{\bk}} \delta_{[r_1+sp]_q,r_2}.
\end{split}
\end{equation}
For well localized Wannier states the dispersion can be approximated by nearest-neighbor hoppings, i.e., $\varepsilon_{\tilde{\bk}} = 2E_0 [\cos(2\pi \tilde{k}_1)+\cos(2\pi \tilde{k}_2)]$, and as a result: 
\begin{equation}\label{eq:harper_hws}
    I_1 \approx  E_0 \left[ 2\cos(2\pi \tilde{k}_2)\delta_{r_1,r_2} + \sum_{s=\pm 1}e^{i2\pi k_1 s} \delta_{[r_1+sp]_q,r_2} \right].
\end{equation}

We compare the above expression to the Harper equation obtained in Ref.~\cite{Hofstadter1976}: 
\begin{equation}
g(m+1) + g(m-1) + 2\cos(\frac{2\pi p}{q}m - 2\pi k_1) g(m) = \frac{E}{E_0} g(m),
\end{equation}
where we have replaced the notation $\nu$ used in Ref.~\cite{Hofstadter1976} with $2\pi k_1$, and $\alpha$ with $p/q$. Performing Fourier series expansion of the wavefunction $g(m)=\sum_{k_2,r_2} e^{i2\pi (k_2+\frac{r_2}{q}) m} g_{k_2,r_2}$ produces the eigen-equation:
\begin{equation}
\begin{split}
& E g_{k_2,r_1} \\
   = & \sum_{r_2}E_0\left[2 \cos(2\pi \tilde{k}_2)\delta_{r_1,r_2} + \sum_{s=\pm 1}e^{i2\pi k_1 s} \delta_{[r_1+sp]_q,r_2} \right] g_{k_2,r_2}.
\end{split}
\end{equation}
The second line is precisely Eq.~(\ref{eq:harper_hws}), explicitly recovering the Harper equation as a limiting case of the finite magnetic field problem when the magnetic length is much longer than the spatial extent of the Wannier orbital. 

{\color{black} In appendix E, we go beyond the nearest neighbor hopping example discussed above, and present a quantitative comparison of the Hofstadter spectra calculated using the Peierls substitution \cite{Peierls1933}, the hWS approach, and the LL approach. As shown in Fig.~\ref{fig:hofstadter_comparison_appendix}, at lower magnetic flux ratios, both Peierls substitution and the hWS approach yield quantitatively accurate Hofstadter spectra. However, as magnetic flux is increased, Peierls substitution approach becomes less accurate than the hWS approach at capturing the qualitative gap structures.}

\section{Summary}
Using the magnetic translation group projection method recently developed in the Ref.~\cite{XW2022} based on the $\bB=0$ hybrid Wannier states (hWS), we re-examined the Hofstadter physics of the $\bB\neq 0$ magnetic subbands corresponding to a subset of energetically isolated, and topologiclaly trivial, Bloch bands. Employing the continuum electron moving in 2D square and triangular lattice potentials as examples, we demonstrated that the method works well up to moderate strengths of magnetic flux per unit cell. Importantly, it naturally bridges the zero field and finite field Hilbert spaces. We also recovered the Harper equation \cite{Hofstadter1976} from the MTG states generated from hWSs and showed that its regime of validity is determined by the ratio of magnetic length to the size of exponentially localized Wannier orbitals.

Although here we only applied this projection method to the non-interacting Hamiltonians, the hWS procedure was shown to be useful in studying interacting Hofstadter problem \cite{XW2022}. This is primarily due to two reasons: (1) At low magnetic fields, the procedure can faithfully construct the correct Hilbert space without having to increase the internal dimensions unlike the conventional LL-based approach, where the upper LL index cutoff increases with decreasing field to achieve numerical convergence. (2) The plane wave nature of the hWSs makes the matrix representation of operators of the form $\mathcal{O}_{\bq}e^{i\bq\cdot\br}$ sparse. By comparison they are dense matrices in the LL wavefunction basis. The method can therefore efficiently address the effects of  Coulomb interactions in a finite magnetic field in the presence of a periodic potential.

\section{Acknowledgements}
X.W. acknowledges financial support from the Gordon and Betty Moore Foundation's EPiQS Initiative Grant GBMF11070. O.V. was supported by NSF Grant No.~DMR-1916958 and is partially funded by the Gordon and Betty Moore Foundation's EPiQS Initiative Grant GBMF11070, National High Magnetic Field Laboratory through NSF Grant No.~DMR-1157490 and the State of Florida.

\bibliographystyle{apsrev4-2}
\bibliography{references}

\begin{thebibliography}{50}%
\makeatletter
\providecommand \@ifxundefined [1]{%
 \@ifx{#1\undefined}
}%
\providecommand \@ifnum [1]{%
 \ifnum #1\expandafter \@firstoftwo
 \else \expandafter \@secondoftwo
 \fi
}%
\providecommand \@ifx [1]{%
 \ifx #1\expandafter \@firstoftwo
 \else \expandafter \@secondoftwo
 \fi
}%
\providecommand \natexlab [1]{#1}%
\providecommand \enquote  [1]{``#1''}%
\providecommand \bibnamefont  [1]{#1}%
\providecommand \bibfnamefont [1]{#1}%
\providecommand \citenamefont [1]{#1}%
\providecommand \href@noop [0]{\@secondoftwo}%
\providecommand \href [0]{\begingroup \@sanitize@url \@href}%
\providecommand \@href[1]{\@@startlink{#1}\@@href}%
\providecommand \@@href[1]{\endgroup#1\@@endlink}%
\providecommand \@sanitize@url [0]{\catcode `\\12\catcode `\$12\catcode
  `\&12\catcode `\#12\catcode `\^12\catcode `\_12\catcode `\%12\relax}%
\providecommand \@@startlink[1]{}%
\providecommand \@@endlink[0]{}%
\providecommand \url  [0]{\begingroup\@sanitize@url \@url }%
\providecommand \@url [1]{\endgroup\@href {#1}{\urlprefix }}%
\providecommand \urlprefix  [0]{URL }%
\providecommand \Eprint [0]{\href }%
\providecommand \doibase [0]{https://doi.org/}%
\providecommand \selectlanguage [0]{\@gobble}%
\providecommand \bibinfo  [0]{\@secondoftwo}%
\providecommand \bibfield  [0]{\@secondoftwo}%
\providecommand \translation [1]{[#1]}%
\providecommand \BibitemOpen [0]{}%
\providecommand \bibitemStop [0]{}%
\providecommand \bibitemNoStop [0]{.\EOS\space}%
\providecommand \EOS [0]{\spacefactor3000\relax}%
\providecommand \BibitemShut  [1]{\csname bibitem#1\endcsname}%
\let\auto@bib@innerbib\@empty
\bibitem [{\citenamefont {Cao}\ \emph {et~al.}(2018{\natexlab{a}})\citenamefont
  {Cao}, \citenamefont {Fatemi}, \citenamefont {Demir}, \citenamefont {Fang},
  \citenamefont {Tomarken}, \citenamefont {Luo}, \citenamefont
  {Sanchez-Yamagishi}, \citenamefont {Watanabe}, \citenamefont {Taniguchi},
  \citenamefont {Kaxiras}, \citenamefont {Ashoori},\ and\ \citenamefont
  {Jarillo-Herrero}}]{Cao2018a}%
  \BibitemOpen
  \bibfield  {author} {\bibinfo {author} {\bibfnamefont {Y.}~\bibnamefont
  {Cao}}, \bibinfo {author} {\bibfnamefont {V.}~\bibnamefont {Fatemi}},
  \bibinfo {author} {\bibfnamefont {A.}~\bibnamefont {Demir}}, \bibinfo
  {author} {\bibfnamefont {S.}~\bibnamefont {Fang}}, \bibinfo {author}
  {\bibfnamefont {S.~L.}\ \bibnamefont {Tomarken}}, \bibinfo {author}
  {\bibfnamefont {J.~Y.}\ \bibnamefont {Luo}}, \bibinfo {author} {\bibfnamefont
  {J.~D.}\ \bibnamefont {Sanchez-Yamagishi}}, \bibinfo {author} {\bibfnamefont
  {K.}~\bibnamefont {Watanabe}}, \bibinfo {author} {\bibfnamefont
  {T.}~\bibnamefont {Taniguchi}}, \bibinfo {author} {\bibfnamefont
  {E.}~\bibnamefont {Kaxiras}}, \bibinfo {author} {\bibfnamefont {R.~C.}\
  \bibnamefont {Ashoori}},\ and\ \bibinfo {author} {\bibfnamefont
  {P.}~\bibnamefont {Jarillo-Herrero}},\ }\href
  {https://doi.org/10.1038/nature26154} {\bibfield  {journal} {\bibinfo
  {journal} {Nature}\ }\textbf {\bibinfo {volume} {556}},\ \bibinfo {pages}
  {80} (\bibinfo {year} {2018}{\natexlab{a}})}\BibitemShut {NoStop}%
\bibitem [{\citenamefont {Cao}\ \emph {et~al.}(2018{\natexlab{b}})\citenamefont
  {Cao}, \citenamefont {Fatemi}, \citenamefont {Fang}, \citenamefont
  {Watanabe}, \citenamefont {Taniguchi}, \citenamefont {Kaxiras},\ and\
  \citenamefont {Jarillo-Herrero}}]{Cao2018b}%
  \BibitemOpen
  \bibfield  {author} {\bibinfo {author} {\bibfnamefont {Y.}~\bibnamefont
  {Cao}}, \bibinfo {author} {\bibfnamefont {V.}~\bibnamefont {Fatemi}},
  \bibinfo {author} {\bibfnamefont {S.}~\bibnamefont {Fang}}, \bibinfo {author}
  {\bibfnamefont {K.}~\bibnamefont {Watanabe}}, \bibinfo {author}
  {\bibfnamefont {T.}~\bibnamefont {Taniguchi}}, \bibinfo {author}
  {\bibfnamefont {E.}~\bibnamefont {Kaxiras}},\ and\ \bibinfo {author}
  {\bibfnamefont {P.}~\bibnamefont {Jarillo-Herrero}},\ }\href
  {https://doi.org/10.1038/nature26160} {\bibfield  {journal} {\bibinfo
  {journal} {Nature}\ }\textbf {\bibinfo {volume} {556}},\ \bibinfo {pages}
  {43} (\bibinfo {year} {2018}{\natexlab{b}})}\BibitemShut {NoStop}%
\bibitem [{\citenamefont {Cao}\ \emph {et~al.}(2020)\citenamefont {Cao},
  \citenamefont {Rodan-Legrain}, \citenamefont {Rubies-Bigorda}, \citenamefont
  {Park}, \citenamefont {Watanabe}, \citenamefont {Taniguchi},\ and\
  \citenamefont {Jarillo-Herrero}}]{Cao2020}%
  \BibitemOpen
  \bibfield  {author} {\bibinfo {author} {\bibfnamefont {Y.}~\bibnamefont
  {Cao}}, \bibinfo {author} {\bibfnamefont {D.}~\bibnamefont {Rodan-Legrain}},
  \bibinfo {author} {\bibfnamefont {O.}~\bibnamefont {Rubies-Bigorda}},
  \bibinfo {author} {\bibfnamefont {J.~M.}\ \bibnamefont {Park}}, \bibinfo
  {author} {\bibfnamefont {K.}~\bibnamefont {Watanabe}}, \bibinfo {author}
  {\bibfnamefont {T.}~\bibnamefont {Taniguchi}},\ and\ \bibinfo {author}
  {\bibfnamefont {P.}~\bibnamefont {Jarillo-Herrero}},\ }\href
  {https://doi.org/10.1038/s41586-020-2260-6} {\bibfield  {journal} {\bibinfo
  {journal} {Nature}\ }\textbf {\bibinfo {volume} {583}},\ \bibinfo {pages}
  {215} (\bibinfo {year} {2020})}\BibitemShut {NoStop}%
\bibitem [{\citenamefont {Tang}\ \emph {et~al.}(2020)\citenamefont {Tang},
  \citenamefont {Li}, \citenamefont {Li}, \citenamefont {Xu}, \citenamefont
  {Liu}, \citenamefont {Barmak}, \citenamefont {Watanabe}, \citenamefont
  {Taniguchi}, \citenamefont {MacDonald}, \citenamefont {Shan},\ and\
  \citenamefont {Mak}}]{Tang2020}%
  \BibitemOpen
  \bibfield  {author} {\bibinfo {author} {\bibfnamefont {Y.}~\bibnamefont
  {Tang}}, \bibinfo {author} {\bibfnamefont {L.}~\bibnamefont {Li}}, \bibinfo
  {author} {\bibfnamefont {T.}~\bibnamefont {Li}}, \bibinfo {author}
  {\bibfnamefont {Y.}~\bibnamefont {Xu}}, \bibinfo {author} {\bibfnamefont
  {S.}~\bibnamefont {Liu}}, \bibinfo {author} {\bibfnamefont {K.}~\bibnamefont
  {Barmak}}, \bibinfo {author} {\bibfnamefont {K.}~\bibnamefont {Watanabe}},
  \bibinfo {author} {\bibfnamefont {T.}~\bibnamefont {Taniguchi}}, \bibinfo
  {author} {\bibfnamefont {A.~H.}\ \bibnamefont {MacDonald}}, \bibinfo {author}
  {\bibfnamefont {J.}~\bibnamefont {Shan}},\ and\ \bibinfo {author}
  {\bibfnamefont {K.~F.}\ \bibnamefont {Mak}},\ }\href
  {https://doi.org/10.1038/s41586-020-2085-3} {\bibfield  {journal} {\bibinfo
  {journal} {Nature}\ }\textbf {\bibinfo {volume} {579}},\ \bibinfo {pages}
  {353} (\bibinfo {year} {2020})}\BibitemShut {NoStop}%
\bibitem [{\citenamefont {Liu}\ \emph {et~al.}(2020)\citenamefont {Liu},
  \citenamefont {Hao}, \citenamefont {Khalaf}, \citenamefont {Lee},
  \citenamefont {Ronen}, \citenamefont {Yoo}, \citenamefont {Haei~Najafabadi},
  \citenamefont {Watanabe}, \citenamefont {Taniguchi}, \citenamefont
  {Vishwanath},\ and\ \citenamefont {Kim}}]{Liu2020}%
  \BibitemOpen
  \bibfield  {author} {\bibinfo {author} {\bibfnamefont {X.}~\bibnamefont
  {Liu}}, \bibinfo {author} {\bibfnamefont {Z.}~\bibnamefont {Hao}}, \bibinfo
  {author} {\bibfnamefont {E.}~\bibnamefont {Khalaf}}, \bibinfo {author}
  {\bibfnamefont {J.~Y.}\ \bibnamefont {Lee}}, \bibinfo {author} {\bibfnamefont
  {Y.}~\bibnamefont {Ronen}}, \bibinfo {author} {\bibfnamefont
  {H.}~\bibnamefont {Yoo}}, \bibinfo {author} {\bibfnamefont {D.}~\bibnamefont
  {Haei~Najafabadi}}, \bibinfo {author} {\bibfnamefont {K.}~\bibnamefont
  {Watanabe}}, \bibinfo {author} {\bibfnamefont {T.}~\bibnamefont {Taniguchi}},
  \bibinfo {author} {\bibfnamefont {A.}~\bibnamefont {Vishwanath}},\ and\
  \bibinfo {author} {\bibfnamefont {P.}~\bibnamefont {Kim}},\ }\href
  {https://doi.org/10.1038/s41586-020-2458-7} {\bibfield  {journal} {\bibinfo
  {journal} {Nature}\ }\textbf {\bibinfo {volume} {583}},\ \bibinfo {pages}
  {221} (\bibinfo {year} {2020})}\BibitemShut {NoStop}%
\bibitem [{\citenamefont {Wang}\ \emph {et~al.}(2020)\citenamefont {Wang},
  \citenamefont {Shih}, \citenamefont {Ghiotto}, \citenamefont {Xian},
  \citenamefont {Rhodes}, \citenamefont {Tan}, \citenamefont {Claassen},
  \citenamefont {Kennes}, \citenamefont {Bai}, \citenamefont {Kim},
  \citenamefont {Watanabe}, \citenamefont {Taniguchi}, \citenamefont {Zhu},
  \citenamefont {Hone}, \citenamefont {Rubio}, \citenamefont {Pasupathy},\ and\
  \citenamefont {Dean}}]{Wang2020}%
  \BibitemOpen
  \bibfield  {author} {\bibinfo {author} {\bibfnamefont {L.}~\bibnamefont
  {Wang}}, \bibinfo {author} {\bibfnamefont {E.-M.}\ \bibnamefont {Shih}},
  \bibinfo {author} {\bibfnamefont {A.}~\bibnamefont {Ghiotto}}, \bibinfo
  {author} {\bibfnamefont {L.}~\bibnamefont {Xian}}, \bibinfo {author}
  {\bibfnamefont {D.~A.}\ \bibnamefont {Rhodes}}, \bibinfo {author}
  {\bibfnamefont {C.}~\bibnamefont {Tan}}, \bibinfo {author} {\bibfnamefont
  {M.}~\bibnamefont {Claassen}}, \bibinfo {author} {\bibfnamefont {D.~M.}\
  \bibnamefont {Kennes}}, \bibinfo {author} {\bibfnamefont {Y.}~\bibnamefont
  {Bai}}, \bibinfo {author} {\bibfnamefont {B.}~\bibnamefont {Kim}}, \bibinfo
  {author} {\bibfnamefont {K.}~\bibnamefont {Watanabe}}, \bibinfo {author}
  {\bibfnamefont {T.}~\bibnamefont {Taniguchi}}, \bibinfo {author}
  {\bibfnamefont {X.}~\bibnamefont {Zhu}}, \bibinfo {author} {\bibfnamefont
  {J.}~\bibnamefont {Hone}}, \bibinfo {author} {\bibfnamefont {A.}~\bibnamefont
  {Rubio}}, \bibinfo {author} {\bibfnamefont {A.~N.}\ \bibnamefont
  {Pasupathy}},\ and\ \bibinfo {author} {\bibfnamefont {C.~R.}\ \bibnamefont
  {Dean}},\ }\href {https://doi.org/10.1038/s41563-020-0708-6} {\bibfield
  {journal} {\bibinfo  {journal} {Nature Materials}\ }\textbf {\bibinfo
  {volume} {19}},\ \bibinfo {pages} {861} (\bibinfo {year} {2020})}\BibitemShut
  {NoStop}%
\bibitem [{\citenamefont {Regan}\ \emph {et~al.}(2020)\citenamefont {Regan},
  \citenamefont {Wang}, \citenamefont {Jin}, \citenamefont {Bakti~Utama},
  \citenamefont {Gao}, \citenamefont {Wei}, \citenamefont {Zhao}, \citenamefont
  {Zhao}, \citenamefont {Zhang}, \citenamefont {Yumigeta}, \citenamefont
  {Blei}, \citenamefont {Carlstr{\"o}m}, \citenamefont {Watanabe},
  \citenamefont {Taniguchi}, \citenamefont {Tongay}, \citenamefont {Crommie},
  \citenamefont {Zettl},\ and\ \citenamefont {Wang}}]{Regan2020}%
  \BibitemOpen
  \bibfield  {author} {\bibinfo {author} {\bibfnamefont {E.~C.}\ \bibnamefont
  {Regan}}, \bibinfo {author} {\bibfnamefont {D.}~\bibnamefont {Wang}},
  \bibinfo {author} {\bibfnamefont {C.}~\bibnamefont {Jin}}, \bibinfo {author}
  {\bibfnamefont {M.~I.}\ \bibnamefont {Bakti~Utama}}, \bibinfo {author}
  {\bibfnamefont {B.}~\bibnamefont {Gao}}, \bibinfo {author} {\bibfnamefont
  {X.}~\bibnamefont {Wei}}, \bibinfo {author} {\bibfnamefont {S.}~\bibnamefont
  {Zhao}}, \bibinfo {author} {\bibfnamefont {W.}~\bibnamefont {Zhao}}, \bibinfo
  {author} {\bibfnamefont {Z.}~\bibnamefont {Zhang}}, \bibinfo {author}
  {\bibfnamefont {K.}~\bibnamefont {Yumigeta}}, \bibinfo {author}
  {\bibfnamefont {M.}~\bibnamefont {Blei}}, \bibinfo {author} {\bibfnamefont
  {J.~D.}\ \bibnamefont {Carlstr{\"o}m}}, \bibinfo {author} {\bibfnamefont
  {K.}~\bibnamefont {Watanabe}}, \bibinfo {author} {\bibfnamefont
  {T.}~\bibnamefont {Taniguchi}}, \bibinfo {author} {\bibfnamefont
  {S.}~\bibnamefont {Tongay}}, \bibinfo {author} {\bibfnamefont
  {M.}~\bibnamefont {Crommie}}, \bibinfo {author} {\bibfnamefont
  {A.}~\bibnamefont {Zettl}},\ and\ \bibinfo {author} {\bibfnamefont
  {F.}~\bibnamefont {Wang}},\ }\href
  {https://doi.org/10.1038/s41586-020-2092-4} {\bibfield  {journal} {\bibinfo
  {journal} {Nature}\ }\textbf {\bibinfo {volume} {579}},\ \bibinfo {pages}
  {359} (\bibinfo {year} {2020})}\BibitemShut {NoStop}%
\bibitem [{\citenamefont {Park}\ \emph {et~al.}(2021)\citenamefont {Park},
  \citenamefont {Cao}, \citenamefont {Watanabe}, \citenamefont {Taniguchi},\
  and\ \citenamefont {Jarillo-Herrero}}]{Park2021}%
  \BibitemOpen
  \bibfield  {author} {\bibinfo {author} {\bibfnamefont {J.~M.}\ \bibnamefont
  {Park}}, \bibinfo {author} {\bibfnamefont {Y.}~\bibnamefont {Cao}}, \bibinfo
  {author} {\bibfnamefont {K.}~\bibnamefont {Watanabe}}, \bibinfo {author}
  {\bibfnamefont {T.}~\bibnamefont {Taniguchi}},\ and\ \bibinfo {author}
  {\bibfnamefont {P.}~\bibnamefont {Jarillo-Herrero}},\ }\href
  {https://doi.org/10.1038/s41586-021-03192-0} {\bibfield  {journal} {\bibinfo
  {journal} {Nature}\ }\textbf {\bibinfo {volume} {590}},\ \bibinfo {pages}
  {249} (\bibinfo {year} {2021})}\BibitemShut {NoStop}%
\bibitem [{\citenamefont {Hao}\ \emph {et~al.}(2021)\citenamefont {Hao},
  \citenamefont {Zimmerman}, \citenamefont {Ledwith}, \citenamefont {Khalaf},
  \citenamefont {Najafabadi}, \citenamefont {Watanabe}, \citenamefont
  {Taniguchi}, \citenamefont {Vishwanath},\ and\ \citenamefont
  {Kim}}]{Hao2021}%
  \BibitemOpen
  \bibfield  {author} {\bibinfo {author} {\bibfnamefont {Z.}~\bibnamefont
  {Hao}}, \bibinfo {author} {\bibfnamefont {A.~M.}\ \bibnamefont {Zimmerman}},
  \bibinfo {author} {\bibfnamefont {P.}~\bibnamefont {Ledwith}}, \bibinfo
  {author} {\bibfnamefont {E.}~\bibnamefont {Khalaf}}, \bibinfo {author}
  {\bibfnamefont {D.~H.}\ \bibnamefont {Najafabadi}}, \bibinfo {author}
  {\bibfnamefont {K.}~\bibnamefont {Watanabe}}, \bibinfo {author}
  {\bibfnamefont {T.}~\bibnamefont {Taniguchi}}, \bibinfo {author}
  {\bibfnamefont {A.}~\bibnamefont {Vishwanath}},\ and\ \bibinfo {author}
  {\bibfnamefont {P.}~\bibnamefont {Kim}},\ }\href
  {https://doi.org/10.1126/science.abg0399} {\bibfield  {journal} {\bibinfo
  {journal} {Science}\ }\textbf {\bibinfo {volume} {371}},\ \bibinfo {pages}
  {1133} (\bibinfo {year} {2021})}\BibitemShut {NoStop}%
\bibitem [{\citenamefont {Kometter}\ \emph {et~al.}(2022)\citenamefont
  {Kometter}, \citenamefont {Yu}, \citenamefont {Devakul}, \citenamefont
  {Reddy}, \citenamefont {Zhang}, \citenamefont {Foutty}, \citenamefont
  {Watanabe}, \citenamefont {Taniguchi}, \citenamefont {Fu},\ and\
  \citenamefont {Feldman}}]{Kometter2022}%
  \BibitemOpen
  \bibfield  {author} {\bibinfo {author} {\bibfnamefont {C.~R.}\ \bibnamefont
  {Kometter}}, \bibinfo {author} {\bibfnamefont {J.}~\bibnamefont {Yu}},
  \bibinfo {author} {\bibfnamefont {T.}~\bibnamefont {Devakul}}, \bibinfo
  {author} {\bibfnamefont {A.~P.}\ \bibnamefont {Reddy}}, \bibinfo {author}
  {\bibfnamefont {Y.}~\bibnamefont {Zhang}}, \bibinfo {author} {\bibfnamefont
  {B.~A.}\ \bibnamefont {Foutty}}, \bibinfo {author} {\bibfnamefont
  {K.}~\bibnamefont {Watanabe}}, \bibinfo {author} {\bibfnamefont
  {T.}~\bibnamefont {Taniguchi}}, \bibinfo {author} {\bibfnamefont
  {L.}~\bibnamefont {Fu}},\ and\ \bibinfo {author} {\bibfnamefont {B.~E.}\
  \bibnamefont {Feldman}},\ }\href {https://doi.org/10.48550/ARXIV.2212.05068}
  {\bibinfo {title} {Hofstadter states and reentrant charge order in a
  semiconductor moiré lattice}} (\bibinfo {year} {2022})\BibitemShut {NoStop}%
\bibitem [{\citenamefont {Yuan}\ and\ \citenamefont {Fu}(2018)}]{Yuan2018}%
  \BibitemOpen
  \bibfield  {author} {\bibinfo {author} {\bibfnamefont {N.~F.~Q.}\
  \bibnamefont {Yuan}}\ and\ \bibinfo {author} {\bibfnamefont {L.}~\bibnamefont
  {Fu}},\ }\href {https://doi.org/10.1103/PhysRevB.98.045103} {\bibfield
  {journal} {\bibinfo  {journal} {Phys. Rev. B}\ }\textbf {\bibinfo {volume}
  {98}},\ \bibinfo {pages} {045103} (\bibinfo {year} {2018})}\BibitemShut
  {NoStop}%
\bibitem [{\citenamefont {Koshino}\ \emph {et~al.}(2018)\citenamefont
  {Koshino}, \citenamefont {Yuan}, \citenamefont {Koretsune}, \citenamefont
  {Ochi}, \citenamefont {Kuroki},\ and\ \citenamefont {Fu}}]{Koshino2018}%
  \BibitemOpen
  \bibfield  {author} {\bibinfo {author} {\bibfnamefont {M.}~\bibnamefont
  {Koshino}}, \bibinfo {author} {\bibfnamefont {N.~F.~Q.}\ \bibnamefont
  {Yuan}}, \bibinfo {author} {\bibfnamefont {T.}~\bibnamefont {Koretsune}},
  \bibinfo {author} {\bibfnamefont {M.}~\bibnamefont {Ochi}}, \bibinfo {author}
  {\bibfnamefont {K.}~\bibnamefont {Kuroki}},\ and\ \bibinfo {author}
  {\bibfnamefont {L.}~\bibnamefont {Fu}},\ }\href
  {https://doi.org/10.1103/PhysRevX.8.031087} {\bibfield  {journal} {\bibinfo
  {journal} {Phys. Rev. X}\ }\textbf {\bibinfo {volume} {8}},\ \bibinfo {pages}
  {031087} (\bibinfo {year} {2018})}\BibitemShut {NoStop}%
\bibitem [{\citenamefont {Kang}\ and\ \citenamefont {Vafek}(2018)}]{Kang2018}%
  \BibitemOpen
  \bibfield  {author} {\bibinfo {author} {\bibfnamefont {J.}~\bibnamefont
  {Kang}}\ and\ \bibinfo {author} {\bibfnamefont {O.}~\bibnamefont {Vafek}},\
  }\href {https://doi.org/10.1103/PhysRevX.8.031088} {\bibfield  {journal}
  {\bibinfo  {journal} {Phys. Rev. X}\ }\textbf {\bibinfo {volume} {8}},\
  \bibinfo {pages} {031088} (\bibinfo {year} {2018})}\BibitemShut {NoStop}%
\bibitem [{\citenamefont {Po}\ \emph {et~al.}(2018{\natexlab{a}})\citenamefont
  {Po}, \citenamefont {Zou}, \citenamefont {Vishwanath},\ and\ \citenamefont
  {Senthil}}]{Po2018}%
  \BibitemOpen
  \bibfield  {author} {\bibinfo {author} {\bibfnamefont {H.~C.}\ \bibnamefont
  {Po}}, \bibinfo {author} {\bibfnamefont {L.}~\bibnamefont {Zou}}, \bibinfo
  {author} {\bibfnamefont {A.}~\bibnamefont {Vishwanath}},\ and\ \bibinfo
  {author} {\bibfnamefont {T.}~\bibnamefont {Senthil}},\ }\href
  {https://doi.org/10.1103/PhysRevX.8.031089} {\bibfield  {journal} {\bibinfo
  {journal} {Phys. Rev. X}\ }\textbf {\bibinfo {volume} {8}},\ \bibinfo {pages}
  {031089} (\bibinfo {year} {2018}{\natexlab{a}})}\BibitemShut {NoStop}%
\bibitem [{\citenamefont {Kang}\ and\ \citenamefont {Vafek}(2019)}]{Kang2019}%
  \BibitemOpen
  \bibfield  {author} {\bibinfo {author} {\bibfnamefont {J.}~\bibnamefont
  {Kang}}\ and\ \bibinfo {author} {\bibfnamefont {O.}~\bibnamefont {Vafek}},\
  }\href {https://doi.org/10.1103/PhysRevLett.122.246401} {\bibfield  {journal}
  {\bibinfo  {journal} {Phys. Rev. Lett.}\ }\textbf {\bibinfo {volume} {122}},\
  \bibinfo {pages} {246401} (\bibinfo {year} {2019})}\BibitemShut {NoStop}%
\bibitem [{\citenamefont {Bernevig}\ \emph
  {et~al.}(2021{\natexlab{a}})\citenamefont {Bernevig}, \citenamefont {Song},
  \citenamefont {Regnault},\ and\ \citenamefont {Lian}}]{TBGI2020}%
  \BibitemOpen
  \bibfield  {author} {\bibinfo {author} {\bibfnamefont {B.~A.}\ \bibnamefont
  {Bernevig}}, \bibinfo {author} {\bibfnamefont {Z.-D.}\ \bibnamefont {Song}},
  \bibinfo {author} {\bibfnamefont {N.}~\bibnamefont {Regnault}},\ and\
  \bibinfo {author} {\bibfnamefont {B.}~\bibnamefont {Lian}},\ }\href
  {https://doi.org/10.1103/PhysRevB.103.205411} {\bibfield  {journal} {\bibinfo
   {journal} {Phys. Rev. B}\ }\textbf {\bibinfo {volume} {103}},\ \bibinfo
  {pages} {205411} (\bibinfo {year} {2021}{\natexlab{a}})}\BibitemShut
  {NoStop}%
\bibitem [{\citenamefont {Song}\ \emph {et~al.}(2021)\citenamefont {Song},
  \citenamefont {Lian}, \citenamefont {Regnault},\ and\ \citenamefont
  {Bernevig}}]{TBGII2020}%
  \BibitemOpen
  \bibfield  {author} {\bibinfo {author} {\bibfnamefont {Z.-D.}\ \bibnamefont
  {Song}}, \bibinfo {author} {\bibfnamefont {B.}~\bibnamefont {Lian}}, \bibinfo
  {author} {\bibfnamefont {N.}~\bibnamefont {Regnault}},\ and\ \bibinfo
  {author} {\bibfnamefont {B.~A.}\ \bibnamefont {Bernevig}},\ }\href
  {https://doi.org/10.1103/PhysRevB.103.205412} {\bibfield  {journal} {\bibinfo
   {journal} {Phys. Rev. B}\ }\textbf {\bibinfo {volume} {103}},\ \bibinfo
  {pages} {205412} (\bibinfo {year} {2021})}\BibitemShut {NoStop}%
\bibitem [{\citenamefont {Bernevig}\ \emph
  {et~al.}(2021{\natexlab{b}})\citenamefont {Bernevig}, \citenamefont {Song},
  \citenamefont {Regnault},\ and\ \citenamefont {Lian}}]{TBGIII2020}%
  \BibitemOpen
  \bibfield  {author} {\bibinfo {author} {\bibfnamefont {B.~A.}\ \bibnamefont
  {Bernevig}}, \bibinfo {author} {\bibfnamefont {Z.-D.}\ \bibnamefont {Song}},
  \bibinfo {author} {\bibfnamefont {N.}~\bibnamefont {Regnault}},\ and\
  \bibinfo {author} {\bibfnamefont {B.}~\bibnamefont {Lian}},\ }\href
  {https://doi.org/10.1103/PhysRevB.103.205413} {\bibfield  {journal} {\bibinfo
   {journal} {Phys. Rev. B}\ }\textbf {\bibinfo {volume} {103}},\ \bibinfo
  {pages} {205413} (\bibinfo {year} {2021}{\natexlab{b}})}\BibitemShut
  {NoStop}%
\bibitem [{\citenamefont {Lian}\ \emph
  {et~al.}(2021{\natexlab{a}})\citenamefont {Lian}, \citenamefont {Song},
  \citenamefont {Regnault}, \citenamefont {Efetov}, \citenamefont {Yazdani},\
  and\ \citenamefont {Bernevig}}]{TBGIV2020}%
  \BibitemOpen
  \bibfield  {author} {\bibinfo {author} {\bibfnamefont {B.}~\bibnamefont
  {Lian}}, \bibinfo {author} {\bibfnamefont {Z.-D.}\ \bibnamefont {Song}},
  \bibinfo {author} {\bibfnamefont {N.}~\bibnamefont {Regnault}}, \bibinfo
  {author} {\bibfnamefont {D.~K.}\ \bibnamefont {Efetov}}, \bibinfo {author}
  {\bibfnamefont {A.}~\bibnamefont {Yazdani}},\ and\ \bibinfo {author}
  {\bibfnamefont {B.~A.}\ \bibnamefont {Bernevig}},\ }\href
  {https://doi.org/10.1103/PhysRevB.103.205414} {\bibfield  {journal} {\bibinfo
   {journal} {Phys. Rev. B}\ }\textbf {\bibinfo {volume} {103}},\ \bibinfo
  {pages} {205414} (\bibinfo {year} {2021}{\natexlab{a}})}\BibitemShut
  {NoStop}%
\bibitem [{\citenamefont {Bernevig}\ \emph
  {et~al.}(2021{\natexlab{c}})\citenamefont {Bernevig}, \citenamefont {Lian},
  \citenamefont {Cowsik}, \citenamefont {Xie}, \citenamefont {Regnault},\ and\
  \citenamefont {Song}}]{TBGV2020}%
  \BibitemOpen
  \bibfield  {author} {\bibinfo {author} {\bibfnamefont {B.~A.}\ \bibnamefont
  {Bernevig}}, \bibinfo {author} {\bibfnamefont {B.}~\bibnamefont {Lian}},
  \bibinfo {author} {\bibfnamefont {A.}~\bibnamefont {Cowsik}}, \bibinfo
  {author} {\bibfnamefont {F.}~\bibnamefont {Xie}}, \bibinfo {author}
  {\bibfnamefont {N.}~\bibnamefont {Regnault}},\ and\ \bibinfo {author}
  {\bibfnamefont {Z.-D.}\ \bibnamefont {Song}},\ }\href
  {https://doi.org/10.1103/PhysRevB.103.205415} {\bibfield  {journal} {\bibinfo
   {journal} {Phys. Rev. B}\ }\textbf {\bibinfo {volume} {103}},\ \bibinfo
  {pages} {205415} (\bibinfo {year} {2021}{\natexlab{c}})}\BibitemShut
  {NoStop}%
\bibitem [{\citenamefont {Balents}\ \emph {et~al.}(2020)\citenamefont
  {Balents}, \citenamefont {Dean}, \citenamefont {Efetov},\ and\ \citenamefont
  {Young}}]{Balents2020}%
  \BibitemOpen
  \bibfield  {author} {\bibinfo {author} {\bibfnamefont {L.}~\bibnamefont
  {Balents}}, \bibinfo {author} {\bibfnamefont {C.~R.}\ \bibnamefont {Dean}},
  \bibinfo {author} {\bibfnamefont {D.~K.}\ \bibnamefont {Efetov}},\ and\
  \bibinfo {author} {\bibfnamefont {A.~F.}\ \bibnamefont {Young}},\ }\href
  {https://doi.org/10.1038/s41567-020-0906-9} {\bibfield  {journal} {\bibinfo
  {journal} {Nature Physics}\ }\textbf {\bibinfo {volume} {16}},\ \bibinfo
  {pages} {725} (\bibinfo {year} {2020})}\BibitemShut {NoStop}%
\bibitem [{\citenamefont {Song}\ and\ \citenamefont
  {Bernevig}(2022)}]{Song2022}%
  \BibitemOpen
  \bibfield  {author} {\bibinfo {author} {\bibfnamefont {Z.-D.}\ \bibnamefont
  {Song}}\ and\ \bibinfo {author} {\bibfnamefont {B.~A.}\ \bibnamefont
  {Bernevig}},\ }\href {https://doi.org/10.1103/PhysRevLett.129.047601}
  {\bibfield  {journal} {\bibinfo  {journal} {Phys. Rev. Lett.}\ }\textbf
  {\bibinfo {volume} {129}},\ \bibinfo {pages} {047601} (\bibinfo {year}
  {2022})}\BibitemShut {NoStop}%
\bibitem [{\citenamefont {Dean}\ \emph {et~al.}(2013)\citenamefont {Dean},
  \citenamefont {Wang}, \citenamefont {Maher}, \citenamefont {Forsythe},
  \citenamefont {Ghahari}, \citenamefont {Gao}, \citenamefont {Katoch},
  \citenamefont {Ishigami}, \citenamefont {Moon}, \citenamefont {Koshino},
  \citenamefont {Taniguchi}, \citenamefont {Watanabe}, \citenamefont {Shepard},
  \citenamefont {Hone},\ and\ \citenamefont {Kim}}]{dean_hofstadters_2013}%
  \BibitemOpen
  \bibfield  {author} {\bibinfo {author} {\bibfnamefont {C.~R.}\ \bibnamefont
  {Dean}}, \bibinfo {author} {\bibfnamefont {L.}~\bibnamefont {Wang}}, \bibinfo
  {author} {\bibfnamefont {P.}~\bibnamefont {Maher}}, \bibinfo {author}
  {\bibfnamefont {C.}~\bibnamefont {Forsythe}}, \bibinfo {author}
  {\bibfnamefont {F.}~\bibnamefont {Ghahari}}, \bibinfo {author} {\bibfnamefont
  {Y.}~\bibnamefont {Gao}}, \bibinfo {author} {\bibfnamefont {J.}~\bibnamefont
  {Katoch}}, \bibinfo {author} {\bibfnamefont {M.}~\bibnamefont {Ishigami}},
  \bibinfo {author} {\bibfnamefont {P.}~\bibnamefont {Moon}}, \bibinfo {author}
  {\bibfnamefont {M.}~\bibnamefont {Koshino}}, \bibinfo {author} {\bibfnamefont
  {T.}~\bibnamefont {Taniguchi}}, \bibinfo {author} {\bibfnamefont
  {K.}~\bibnamefont {Watanabe}}, \bibinfo {author} {\bibfnamefont {K.~L.}\
  \bibnamefont {Shepard}}, \bibinfo {author} {\bibfnamefont {J.}~\bibnamefont
  {Hone}},\ and\ \bibinfo {author} {\bibfnamefont {P.}~\bibnamefont {Kim}},\
  }\href {https://doi.org/10.1038/nature12186} {\bibfield  {journal} {\bibinfo
  {journal} {Nature}\ }\textbf {\bibinfo {volume} {497}},\ \bibinfo {pages}
  {598} (\bibinfo {year} {2013})}\BibitemShut {NoStop}%
\bibitem [{\citenamefont {Lu}\ \emph {et~al.}(2019)\citenamefont {Lu},
  \citenamefont {Stepanov}, \citenamefont {Yang}, \citenamefont {Xie},
  \citenamefont {Aamir}, \citenamefont {Das}, \citenamefont {Urgell},
  \citenamefont {Watanabe}, \citenamefont {Taniguchi}, \citenamefont {Zhang},
  \citenamefont {Bachtold}, \citenamefont {MacDonald},\ and\ \citenamefont
  {Efetov}}]{Lu2019}%
  \BibitemOpen
  \bibfield  {author} {\bibinfo {author} {\bibfnamefont {X.}~\bibnamefont
  {Lu}}, \bibinfo {author} {\bibfnamefont {P.}~\bibnamefont {Stepanov}},
  \bibinfo {author} {\bibfnamefont {W.}~\bibnamefont {Yang}}, \bibinfo {author}
  {\bibfnamefont {M.}~\bibnamefont {Xie}}, \bibinfo {author} {\bibfnamefont
  {M.~A.}\ \bibnamefont {Aamir}}, \bibinfo {author} {\bibfnamefont
  {I.}~\bibnamefont {Das}}, \bibinfo {author} {\bibfnamefont {C.}~\bibnamefont
  {Urgell}}, \bibinfo {author} {\bibfnamefont {K.}~\bibnamefont {Watanabe}},
  \bibinfo {author} {\bibfnamefont {T.}~\bibnamefont {Taniguchi}}, \bibinfo
  {author} {\bibfnamefont {G.}~\bibnamefont {Zhang}}, \bibinfo {author}
  {\bibfnamefont {A.}~\bibnamefont {Bachtold}}, \bibinfo {author}
  {\bibfnamefont {A.~H.}\ \bibnamefont {MacDonald}},\ and\ \bibinfo {author}
  {\bibfnamefont {D.~K.}\ \bibnamefont {Efetov}},\ }\href
  {https://doi.org/10.1038/s41586-019-1695-0} {\bibfield  {journal} {\bibinfo
  {journal} {Nature}\ }\textbf {\bibinfo {volume} {574}},\ \bibinfo {pages}
  {653} (\bibinfo {year} {2019})}\BibitemShut {NoStop}%
\bibitem [{\citenamefont {Yankowitz}\ \emph {et~al.}(2019)\citenamefont
  {Yankowitz}, \citenamefont {Chen}, \citenamefont {Polshyn}, \citenamefont
  {Zhang}, \citenamefont {Watanabe}, \citenamefont {Taniguchi}, \citenamefont
  {Graf}, \citenamefont {Young},\ and\ \citenamefont {Dean}}]{Yankowitz2019}%
  \BibitemOpen
  \bibfield  {author} {\bibinfo {author} {\bibfnamefont {M.}~\bibnamefont
  {Yankowitz}}, \bibinfo {author} {\bibfnamefont {S.}~\bibnamefont {Chen}},
  \bibinfo {author} {\bibfnamefont {H.}~\bibnamefont {Polshyn}}, \bibinfo
  {author} {\bibfnamefont {Y.}~\bibnamefont {Zhang}}, \bibinfo {author}
  {\bibfnamefont {K.}~\bibnamefont {Watanabe}}, \bibinfo {author}
  {\bibfnamefont {T.}~\bibnamefont {Taniguchi}}, \bibinfo {author}
  {\bibfnamefont {D.}~\bibnamefont {Graf}}, \bibinfo {author} {\bibfnamefont
  {A.~F.}\ \bibnamefont {Young}},\ and\ \bibinfo {author} {\bibfnamefont
  {C.~R.}\ \bibnamefont {Dean}},\ }\href
  {https://doi.org/10.1126/science.aav1910} {\bibfield  {journal} {\bibinfo
  {journal} {Science}\ }\textbf {\bibinfo {volume} {363}},\ \bibinfo {pages}
  {1059} (\bibinfo {year} {2019})}\BibitemShut {NoStop}%
\bibitem [{\citenamefont {Sharpe}\ \emph {et~al.}(2019)\citenamefont {Sharpe},
  \citenamefont {Fox}, \citenamefont {Barnard}, \citenamefont {Finney},
  \citenamefont {Watanabe}, \citenamefont {Taniguchi}, \citenamefont
  {Kastner},\ and\ \citenamefont {Goldhaber-Gordon}}]{Sharpe2019}%
  \BibitemOpen
  \bibfield  {author} {\bibinfo {author} {\bibfnamefont {A.~L.}\ \bibnamefont
  {Sharpe}}, \bibinfo {author} {\bibfnamefont {E.~J.}\ \bibnamefont {Fox}},
  \bibinfo {author} {\bibfnamefont {A.~W.}\ \bibnamefont {Barnard}}, \bibinfo
  {author} {\bibfnamefont {J.}~\bibnamefont {Finney}}, \bibinfo {author}
  {\bibfnamefont {K.}~\bibnamefont {Watanabe}}, \bibinfo {author}
  {\bibfnamefont {T.}~\bibnamefont {Taniguchi}}, \bibinfo {author}
  {\bibfnamefont {M.~A.}\ \bibnamefont {Kastner}},\ and\ \bibinfo {author}
  {\bibfnamefont {D.}~\bibnamefont {Goldhaber-Gordon}},\ }\href
  {https://doi.org/10.1126/science.aaw3780} {\bibfield  {journal} {\bibinfo
  {journal} {Science}\ }\textbf {\bibinfo {volume} {365}},\ \bibinfo {pages}
  {605} (\bibinfo {year} {2019})}\BibitemShut {NoStop}%
\bibitem [{\citenamefont {Nuckolls}\ \emph {et~al.}(2020)\citenamefont
  {Nuckolls}, \citenamefont {Oh}, \citenamefont {Wong}, \citenamefont {Lian},
  \citenamefont {Watanabe}, \citenamefont {Taniguchi}, \citenamefont
  {Bernevig},\ and\ \citenamefont {Yazdani}}]{Nuckolls2020}%
  \BibitemOpen
  \bibfield  {author} {\bibinfo {author} {\bibfnamefont {K.~P.}\ \bibnamefont
  {Nuckolls}}, \bibinfo {author} {\bibfnamefont {M.}~\bibnamefont {Oh}},
  \bibinfo {author} {\bibfnamefont {D.}~\bibnamefont {Wong}}, \bibinfo {author}
  {\bibfnamefont {B.}~\bibnamefont {Lian}}, \bibinfo {author} {\bibfnamefont
  {K.}~\bibnamefont {Watanabe}}, \bibinfo {author} {\bibfnamefont
  {T.}~\bibnamefont {Taniguchi}}, \bibinfo {author} {\bibfnamefont {B.~A.}\
  \bibnamefont {Bernevig}},\ and\ \bibinfo {author} {\bibfnamefont
  {A.}~\bibnamefont {Yazdani}},\ }\href
  {https://doi.org/10.1038/s41586-020-3028-8} {\bibfield  {journal} {\bibinfo
  {journal} {Nature}\ }\textbf {\bibinfo {volume} {588}},\ \bibinfo {pages}
  {610} (\bibinfo {year} {2020})}\BibitemShut {NoStop}%
\bibitem [{\citenamefont {Pierce}\ \emph {et~al.}(2021)\citenamefont {Pierce},
  \citenamefont {Xie}, \citenamefont {Park}, \citenamefont {Khalaf},
  \citenamefont {Lee}, \citenamefont {Cao}, \citenamefont {Parker},
  \citenamefont {Forrester}, \citenamefont {Chen}, \citenamefont {Watanabe},
  \citenamefont {Taniguchi}, \citenamefont {Vishwanath}, \citenamefont
  {Jarillo-Herrero},\ and\ \citenamefont {Yacoby}}]{Pierce2021}%
  \BibitemOpen
  \bibfield  {author} {\bibinfo {author} {\bibfnamefont {A.~T.}\ \bibnamefont
  {Pierce}}, \bibinfo {author} {\bibfnamefont {Y.}~\bibnamefont {Xie}},
  \bibinfo {author} {\bibfnamefont {J.~M.}\ \bibnamefont {Park}}, \bibinfo
  {author} {\bibfnamefont {E.}~\bibnamefont {Khalaf}}, \bibinfo {author}
  {\bibfnamefont {S.~H.}\ \bibnamefont {Lee}}, \bibinfo {author} {\bibfnamefont
  {Y.}~\bibnamefont {Cao}}, \bibinfo {author} {\bibfnamefont {D.~E.}\
  \bibnamefont {Parker}}, \bibinfo {author} {\bibfnamefont {P.~R.}\
  \bibnamefont {Forrester}}, \bibinfo {author} {\bibfnamefont {S.}~\bibnamefont
  {Chen}}, \bibinfo {author} {\bibfnamefont {K.}~\bibnamefont {Watanabe}},
  \bibinfo {author} {\bibfnamefont {T.}~\bibnamefont {Taniguchi}}, \bibinfo
  {author} {\bibfnamefont {A.}~\bibnamefont {Vishwanath}}, \bibinfo {author}
  {\bibfnamefont {P.}~\bibnamefont {Jarillo-Herrero}},\ and\ \bibinfo {author}
  {\bibfnamefont {A.}~\bibnamefont {Yacoby}},\ }\href
  {https://doi.org/10.1038/s41567-021-01347-4} {\bibfield  {journal} {\bibinfo
  {journal} {Nature Physics}\ }\textbf {\bibinfo {volume} {17}},\ \bibinfo
  {pages} {1210} (\bibinfo {year} {2021})}\BibitemShut {NoStop}%
\bibitem [{\citenamefont {Saito}\ \emph {et~al.}(2021)\citenamefont {Saito},
  \citenamefont {Ge}, \citenamefont {Rademaker}, \citenamefont {Watanabe},
  \citenamefont {Taniguchi}, \citenamefont {Abanin},\ and\ \citenamefont
  {Young}}]{Saito2021}%
  \BibitemOpen
  \bibfield  {author} {\bibinfo {author} {\bibfnamefont {Y.}~\bibnamefont
  {Saito}}, \bibinfo {author} {\bibfnamefont {J.}~\bibnamefont {Ge}}, \bibinfo
  {author} {\bibfnamefont {L.}~\bibnamefont {Rademaker}}, \bibinfo {author}
  {\bibfnamefont {K.}~\bibnamefont {Watanabe}}, \bibinfo {author}
  {\bibfnamefont {T.}~\bibnamefont {Taniguchi}}, \bibinfo {author}
  {\bibfnamefont {D.~A.}\ \bibnamefont {Abanin}},\ and\ \bibinfo {author}
  {\bibfnamefont {A.~F.}\ \bibnamefont {Young}},\ }\href
  {https://doi.org/10.1038/s41567-020-01129-4} {\bibfield  {journal} {\bibinfo
  {journal} {Nature Physics}\ }\textbf {\bibinfo {volume} {17}},\ \bibinfo
  {pages} {478} (\bibinfo {year} {2021})}\BibitemShut {NoStop}%
\bibitem [{\citenamefont {Wu}\ \emph {et~al.}(2021)\citenamefont {Wu},
  \citenamefont {Zhang}, \citenamefont {Watanabe}, \citenamefont {Taniguchi},\
  and\ \citenamefont {Andrei}}]{Wu2021}%
  \BibitemOpen
  \bibfield  {author} {\bibinfo {author} {\bibfnamefont {S.}~\bibnamefont
  {Wu}}, \bibinfo {author} {\bibfnamefont {Z.}~\bibnamefont {Zhang}}, \bibinfo
  {author} {\bibfnamefont {K.}~\bibnamefont {Watanabe}}, \bibinfo {author}
  {\bibfnamefont {T.}~\bibnamefont {Taniguchi}},\ and\ \bibinfo {author}
  {\bibfnamefont {E.~Y.}\ \bibnamefont {Andrei}},\ }\href
  {https://doi.org/10.1038/s41563-020-00911-2} {\bibfield  {journal} {\bibinfo
  {journal} {Nature Materials}\ }\textbf {\bibinfo {volume} {20}},\ \bibinfo
  {pages} {488} (\bibinfo {year} {2021})}\BibitemShut {NoStop}%
\bibitem [{\citenamefont {Finney}\ \emph {et~al.}(2022)\citenamefont {Finney},
  \citenamefont {Sharpe}, \citenamefont {Fox}, \citenamefont {Hsueh},
  \citenamefont {Parker}, \citenamefont {Yankowitz}, \citenamefont {Chen},
  \citenamefont {Watanabe}, \citenamefont {Taniguchi}, \citenamefont {Dean},
  \citenamefont {Vishwanath}, \citenamefont {Kastner},\ and\ \citenamefont
  {Goldhaber-Gordon}}]{Finney2022}%
  \BibitemOpen
  \bibfield  {author} {\bibinfo {author} {\bibfnamefont {J.}~\bibnamefont
  {Finney}}, \bibinfo {author} {\bibfnamefont {A.~L.}\ \bibnamefont {Sharpe}},
  \bibinfo {author} {\bibfnamefont {E.~J.}\ \bibnamefont {Fox}}, \bibinfo
  {author} {\bibfnamefont {C.~L.}\ \bibnamefont {Hsueh}}, \bibinfo {author}
  {\bibfnamefont {D.~E.}\ \bibnamefont {Parker}}, \bibinfo {author}
  {\bibfnamefont {M.}~\bibnamefont {Yankowitz}}, \bibinfo {author}
  {\bibfnamefont {S.}~\bibnamefont {Chen}}, \bibinfo {author} {\bibfnamefont
  {K.}~\bibnamefont {Watanabe}}, \bibinfo {author} {\bibfnamefont
  {T.}~\bibnamefont {Taniguchi}}, \bibinfo {author} {\bibfnamefont {C.~R.}\
  \bibnamefont {Dean}}, \bibinfo {author} {\bibfnamefont {A.}~\bibnamefont
  {Vishwanath}}, \bibinfo {author} {\bibfnamefont {M.~A.}\ \bibnamefont
  {Kastner}},\ and\ \bibinfo {author} {\bibfnamefont {D.}~\bibnamefont
  {Goldhaber-Gordon}},\ }\href {https://doi.org/10.1073/pnas.2118482119}
  {\bibfield  {journal} {\bibinfo  {journal} {Proceedings of the National
  Academy of Sciences}\ }\textbf {\bibinfo {volume} {119}},\ \bibinfo {pages}
  {e2118482119} (\bibinfo {year} {2022})}\BibitemShut {NoStop}%
\bibitem [{\citenamefont {Yu}\ \emph {et~al.}(2022)\citenamefont {Yu},
  \citenamefont {Foutty}, \citenamefont {Han}, \citenamefont {Barber},
  \citenamefont {Schattner}, \citenamefont {Watanabe}, \citenamefont
  {Taniguchi}, \citenamefont {Phillips}, \citenamefont {Shen}, \citenamefont
  {Kivelson},\ and\ \citenamefont {Feldman}}]{Yu2022}%
  \BibitemOpen
  \bibfield  {author} {\bibinfo {author} {\bibfnamefont {J.}~\bibnamefont
  {Yu}}, \bibinfo {author} {\bibfnamefont {B.~A.}\ \bibnamefont {Foutty}},
  \bibinfo {author} {\bibfnamefont {Z.}~\bibnamefont {Han}}, \bibinfo {author}
  {\bibfnamefont {M.~E.}\ \bibnamefont {Barber}}, \bibinfo {author}
  {\bibfnamefont {Y.}~\bibnamefont {Schattner}}, \bibinfo {author}
  {\bibfnamefont {K.}~\bibnamefont {Watanabe}}, \bibinfo {author}
  {\bibfnamefont {T.}~\bibnamefont {Taniguchi}}, \bibinfo {author}
  {\bibfnamefont {P.}~\bibnamefont {Phillips}}, \bibinfo {author}
  {\bibfnamefont {Z.-X.}\ \bibnamefont {Shen}}, \bibinfo {author}
  {\bibfnamefont {S.~A.}\ \bibnamefont {Kivelson}},\ and\ \bibinfo {author}
  {\bibfnamefont {B.~E.}\ \bibnamefont {Feldman}},\ }\href
  {https://doi.org/10.1038/s41567-022-01589-w} {\bibfield  {journal} {\bibinfo
  {journal} {Nature Physics}\ }\textbf {\bibinfo {volume} {18}},\ \bibinfo
  {pages} {825} (\bibinfo {year} {2022})}\BibitemShut {NoStop}%
\bibitem [{\citenamefont {Hejazi}\ \emph {et~al.}(2019)\citenamefont {Hejazi},
  \citenamefont {Liu},\ and\ \citenamefont {Balents}}]{Hejazi2019}%
  \BibitemOpen
  \bibfield  {author} {\bibinfo {author} {\bibfnamefont {K.}~\bibnamefont
  {Hejazi}}, \bibinfo {author} {\bibfnamefont {C.}~\bibnamefont {Liu}},\ and\
  \bibinfo {author} {\bibfnamefont {L.}~\bibnamefont {Balents}},\ }\href
  {https://doi.org/10.1103/PhysRevB.100.035115} {\bibfield  {journal} {\bibinfo
   {journal} {Phys. Rev. B}\ }\textbf {\bibinfo {volume} {100}},\ \bibinfo
  {pages} {035115} (\bibinfo {year} {2019})}\BibitemShut {NoStop}%
\bibitem [{\citenamefont {Zhang}\ \emph {et~al.}(2019)\citenamefont {Zhang},
  \citenamefont {Po},\ and\ \citenamefont {Senthil}}]{Yahui2019}%
  \BibitemOpen
  \bibfield  {author} {\bibinfo {author} {\bibfnamefont {Y.-H.}\ \bibnamefont
  {Zhang}}, \bibinfo {author} {\bibfnamefont {H.~C.}\ \bibnamefont {Po}},\ and\
  \bibinfo {author} {\bibfnamefont {T.}~\bibnamefont {Senthil}},\ }\href
  {https://doi.org/10.1103/PhysRevB.100.125104} {\bibfield  {journal} {\bibinfo
   {journal} {Phys. Rev. B}\ }\textbf {\bibinfo {volume} {100}},\ \bibinfo
  {pages} {125104} (\bibinfo {year} {2019})}\BibitemShut {NoStop}%
\bibitem [{\citenamefont {Lian}\ \emph
  {et~al.}(2021{\natexlab{b}})\citenamefont {Lian}, \citenamefont {Xie},\ and\
  \citenamefont {Bernevig}}]{Biao2021LL}%
  \BibitemOpen
  \bibfield  {author} {\bibinfo {author} {\bibfnamefont {B.}~\bibnamefont
  {Lian}}, \bibinfo {author} {\bibfnamefont {F.}~\bibnamefont {Xie}},\ and\
  \bibinfo {author} {\bibfnamefont {B.~A.}\ \bibnamefont {Bernevig}},\ }\href
  {https://doi.org/10.1103/PhysRevB.103.L161405} {\bibfield  {journal}
  {\bibinfo  {journal} {Phys. Rev. B}\ }\textbf {\bibinfo {volume} {103}},\
  \bibinfo {pages} {L161405} (\bibinfo {year}
  {2021}{\natexlab{b}})}\BibitemShut {NoStop}%
\bibitem [{\citenamefont {Herzog-Arbeitman}\ \emph
  {et~al.}(2022{\natexlab{a}})\citenamefont {Herzog-Arbeitman}, \citenamefont
  {Chew}, \citenamefont {Efetov},\ and\ \citenamefont {Bernevig}}]{Jonah2021}%
  \BibitemOpen
  \bibfield  {author} {\bibinfo {author} {\bibfnamefont {J.}~\bibnamefont
  {Herzog-Arbeitman}}, \bibinfo {author} {\bibfnamefont {A.}~\bibnamefont
  {Chew}}, \bibinfo {author} {\bibfnamefont {D.~K.}\ \bibnamefont {Efetov}},\
  and\ \bibinfo {author} {\bibfnamefont {B.~A.}\ \bibnamefont {Bernevig}},\
  }\href {https://doi.org/10.1103/PhysRevLett.129.076401} {\bibfield  {journal}
  {\bibinfo  {journal} {Phys. Rev. Lett.}\ }\textbf {\bibinfo {volume} {129}},\
  \bibinfo {pages} {076401} (\bibinfo {year} {2022}{\natexlab{a}})}\BibitemShut
  {NoStop}%
\bibitem [{\citenamefont {Parker}\ \emph {et~al.}(2021)\citenamefont {Parker},
  \citenamefont {Ledwith}, \citenamefont {Khalaf}, \citenamefont {Soejima},
  \citenamefont {Hauschild}, \citenamefont {Xie}, \citenamefont {Pierce},
  \citenamefont {Zaletel}, \citenamefont {Yacoby},\ and\ \citenamefont
  {Vishwanath}}]{Parker2021b}%
  \BibitemOpen
  \bibfield  {author} {\bibinfo {author} {\bibfnamefont {D.}~\bibnamefont
  {Parker}}, \bibinfo {author} {\bibfnamefont {P.}~\bibnamefont {Ledwith}},
  \bibinfo {author} {\bibfnamefont {E.}~\bibnamefont {Khalaf}}, \bibinfo
  {author} {\bibfnamefont {T.}~\bibnamefont {Soejima}}, \bibinfo {author}
  {\bibfnamefont {J.}~\bibnamefont {Hauschild}}, \bibinfo {author}
  {\bibfnamefont {Y.}~\bibnamefont {Xie}}, \bibinfo {author} {\bibfnamefont
  {A.}~\bibnamefont {Pierce}}, \bibinfo {author} {\bibfnamefont {M.~P.}\
  \bibnamefont {Zaletel}}, \bibinfo {author} {\bibfnamefont {A.}~\bibnamefont
  {Yacoby}},\ and\ \bibinfo {author} {\bibfnamefont {A.}~\bibnamefont
  {Vishwanath}},\ }\Eprint {https://arxiv.org/abs/arXiv:2112.13837}
  {arXiv:2112.13837}  (\bibinfo {year} {2021})\BibitemShut {NoStop}%
\bibitem [{\citenamefont {Herzog-Arbeitman}\ \emph
  {et~al.}(2022{\natexlab{b}})\citenamefont {Herzog-Arbeitman}, \citenamefont
  {Chew},\ and\ \citenamefont {Bernevig}}]{Jonah2022}%
  \BibitemOpen
  \bibfield  {author} {\bibinfo {author} {\bibfnamefont {J.}~\bibnamefont
  {Herzog-Arbeitman}}, \bibinfo {author} {\bibfnamefont {A.}~\bibnamefont
  {Chew}},\ and\ \bibinfo {author} {\bibfnamefont {B.~A.}\ \bibnamefont
  {Bernevig}},\ }\href {https://doi.org/10.1103/PhysRevB.106.085140} {\bibfield
   {journal} {\bibinfo  {journal} {Phys. Rev. B}\ }\textbf {\bibinfo {volume}
  {106}},\ \bibinfo {pages} {085140} (\bibinfo {year}
  {2022}{\natexlab{b}})}\BibitemShut {NoStop}%
\bibitem [{\citenamefont {Peierls}(1933)}]{Peierls1933}%
  \BibitemOpen
  \bibfield  {author} {\bibinfo {author} {\bibfnamefont {R.}~\bibnamefont
  {Peierls}},\ }\href {https://doi.org/10.1007/BF01342591} {\bibfield
  {journal} {\bibinfo  {journal} {Zeitschrift f{\"u}r Physik}\ }\textbf
  {\bibinfo {volume} {80}},\ \bibinfo {pages} {763} (\bibinfo {year}
  {1933})}\BibitemShut {NoStop}%
\bibitem [{\citenamefont {Luttinger}(1951)}]{Luttinger1951}%
  \BibitemOpen
  \bibfield  {author} {\bibinfo {author} {\bibfnamefont {J.~M.}\ \bibnamefont
  {Luttinger}},\ }\href {https://doi.org/10.1103/PhysRev.84.814} {\bibfield
  {journal} {\bibinfo  {journal} {Phys. Rev.}\ }\textbf {\bibinfo {volume}
  {84}},\ \bibinfo {pages} {814} (\bibinfo {year} {1951})}\BibitemShut
  {NoStop}%
\bibitem [{\citenamefont {Harper}(1955)}]{Harper1955}%
  \BibitemOpen
  \bibfield  {author} {\bibinfo {author} {\bibfnamefont {P.~G.}\ \bibnamefont
  {Harper}},\ }\href {https://doi.org/10.1088/0370-1298/68/10/304} {\bibfield
  {journal} {\bibinfo  {journal} {Proceedings of the Physical Society. Section
  A}\ }\textbf {\bibinfo {volume} {68}},\ \bibinfo {pages} {874} (\bibinfo
  {year} {1955})}\BibitemShut {NoStop}%
\bibitem [{\citenamefont {Hofstadter}(1976)}]{Hofstadter1976}%
  \BibitemOpen
  \bibfield  {author} {\bibinfo {author} {\bibfnamefont {D.~R.}\ \bibnamefont
  {Hofstadter}},\ }\href {https://doi.org/10.1103/PhysRevB.14.2239} {\bibfield
  {journal} {\bibinfo  {journal} {Phys. Rev. B}\ }\textbf {\bibinfo {volume}
  {14}},\ \bibinfo {pages} {2239} (\bibinfo {year} {1976})}\BibitemShut
  {NoStop}%
\bibitem [{\citenamefont {Soluyanov}\ and\ \citenamefont
  {Vanderbilt}(2011)}]{Soluyanov2011}%
  \BibitemOpen
  \bibfield  {author} {\bibinfo {author} {\bibfnamefont {A.~A.}\ \bibnamefont
  {Soluyanov}}\ and\ \bibinfo {author} {\bibfnamefont {D.}~\bibnamefont
  {Vanderbilt}},\ }\href {https://doi.org/10.1103/PhysRevB.83.035108}
  {\bibfield  {journal} {\bibinfo  {journal} {Phys. Rev. B}\ }\textbf {\bibinfo
  {volume} {83}},\ \bibinfo {pages} {035108} (\bibinfo {year}
  {2011})}\BibitemShut {NoStop}%
\bibitem [{\citenamefont {Po}\ \emph {et~al.}(2018{\natexlab{b}})\citenamefont
  {Po}, \citenamefont {Watanabe},\ and\ \citenamefont {Vishwanath}}]{Po2018b}%
  \BibitemOpen
  \bibfield  {author} {\bibinfo {author} {\bibfnamefont {H.~C.}\ \bibnamefont
  {Po}}, \bibinfo {author} {\bibfnamefont {H.}~\bibnamefont {Watanabe}},\ and\
  \bibinfo {author} {\bibfnamefont {A.}~\bibnamefont {Vishwanath}},\ }\href
  {https://doi.org/10.1103/PhysRevLett.121.126402} {\bibfield  {journal}
  {\bibinfo  {journal} {Phys. Rev. Lett.}\ }\textbf {\bibinfo {volume} {121}},\
  \bibinfo {pages} {126402} (\bibinfo {year} {2018}{\natexlab{b}})}\BibitemShut
  {NoStop}%
\bibitem [{\citenamefont {Wang}\ and\ \citenamefont {Vafek}(2020)}]{XW2020}%
  \BibitemOpen
  \bibfield  {author} {\bibinfo {author} {\bibfnamefont {X.}~\bibnamefont
  {Wang}}\ and\ \bibinfo {author} {\bibfnamefont {O.}~\bibnamefont {Vafek}},\
  }\href {https://doi.org/10.1103/PhysRevB.102.075142} {\bibfield  {journal}
  {\bibinfo  {journal} {Phys. Rev. B}\ }\textbf {\bibinfo {volume} {102}},\
  \bibinfo {pages} {075142} (\bibinfo {year} {2020})}\BibitemShut {NoStop}%
\bibitem [{\citenamefont {Wang}\ and\ \citenamefont {Vafek}(2022)}]{XW2022}%
  \BibitemOpen
  \bibfield  {author} {\bibinfo {author} {\bibfnamefont {X.}~\bibnamefont
  {Wang}}\ and\ \bibinfo {author} {\bibfnamefont {O.}~\bibnamefont {Vafek}},\
  }\href {https://doi.org/10.1103/PhysRevB.106.L121111} {\bibfield  {journal}
  {\bibinfo  {journal} {Phys. Rev. B}\ }\textbf {\bibinfo {volume} {106}},\
  \bibinfo {pages} {L121111} (\bibinfo {year} {2022})}\BibitemShut {NoStop}%
\bibitem [{\citenamefont {Sgiarovello}\ \emph {et~al.}(2001)\citenamefont
  {Sgiarovello}, \citenamefont {Peressi},\ and\ \citenamefont
  {Resta}}]{Resta2001}%
  \BibitemOpen
  \bibfield  {author} {\bibinfo {author} {\bibfnamefont {C.}~\bibnamefont
  {Sgiarovello}}, \bibinfo {author} {\bibfnamefont {M.}~\bibnamefont
  {Peressi}},\ and\ \bibinfo {author} {\bibfnamefont {R.}~\bibnamefont
  {Resta}},\ }\href {https://doi.org/10.1103/PhysRevB.64.115202} {\bibfield
  {journal} {\bibinfo  {journal} {Phys. Rev. B}\ }\textbf {\bibinfo {volume}
  {64}},\ \bibinfo {pages} {115202} (\bibinfo {year} {2001})}\BibitemShut
  {NoStop}%
\bibitem [{\citenamefont {Yu}\ \emph {et~al.}(2011)\citenamefont {Yu},
  \citenamefont {Qi}, \citenamefont {Bernevig}, \citenamefont {Fang},\ and\
  \citenamefont {Dai}}]{Rui2011}%
  \BibitemOpen
  \bibfield  {author} {\bibinfo {author} {\bibfnamefont {R.}~\bibnamefont
  {Yu}}, \bibinfo {author} {\bibfnamefont {X.~L.}\ \bibnamefont {Qi}}, \bibinfo
  {author} {\bibfnamefont {A.}~\bibnamefont {Bernevig}}, \bibinfo {author}
  {\bibfnamefont {Z.}~\bibnamefont {Fang}},\ and\ \bibinfo {author}
  {\bibfnamefont {X.}~\bibnamefont {Dai}},\ }\href
  {https://doi.org/10.1103/PhysRevB.84.075119} {\bibfield  {journal} {\bibinfo
  {journal} {Phys. Rev. B}\ }\textbf {\bibinfo {volume} {84}},\ \bibinfo
  {pages} {075119} (\bibinfo {year} {2011})}\BibitemShut {NoStop}%
\bibitem [{\citenamefont {Marzari}\ \emph {et~al.}(2012)\citenamefont
  {Marzari}, \citenamefont {Mostofi}, \citenamefont {Yates}, \citenamefont
  {Souza},\ and\ \citenamefont {Vanderbilt}}]{Mazari2012}%
  \BibitemOpen
  \bibfield  {author} {\bibinfo {author} {\bibfnamefont {N.}~\bibnamefont
  {Marzari}}, \bibinfo {author} {\bibfnamefont {A.~A.}\ \bibnamefont
  {Mostofi}}, \bibinfo {author} {\bibfnamefont {J.~R.}\ \bibnamefont {Yates}},
  \bibinfo {author} {\bibfnamefont {I.}~\bibnamefont {Souza}},\ and\ \bibinfo
  {author} {\bibfnamefont {D.}~\bibnamefont {Vanderbilt}},\ }\href
  {https://doi.org/10.1103/RevModPhys.84.1419} {\bibfield  {journal} {\bibinfo
  {journal} {Rev. Mod. Phys.}\ }\textbf {\bibinfo {volume} {84}},\ \bibinfo
  {pages} {1419} (\bibinfo {year} {2012})}\BibitemShut {NoStop}%
\bibitem [{\citenamefont {Kang}\ and\ \citenamefont {Vafek}(2020)}]{Kang2020a}%
  \BibitemOpen
  \bibfield  {author} {\bibinfo {author} {\bibfnamefont {J.}~\bibnamefont
  {Kang}}\ and\ \bibinfo {author} {\bibfnamefont {O.}~\bibnamefont {Vafek}},\
  }\href {https://doi.org/10.1103/PhysRevB.102.035161} {\bibfield  {journal}
  {\bibinfo  {journal} {Phys. Rev. B}\ }\textbf {\bibinfo {volume} {102}},\
  \bibinfo {pages} {035161} (\bibinfo {year} {2020})}\BibitemShut {NoStop}%
\end{thebibliography}%

\begin{widetext}
\appendix 
\renewcommand\thefigure{\thesection.\arabic{figure}}    
\setcounter{figure}{0}    

\section{Landau gauge magnetic translation group identities}
Here we derive an identity for $\hat{t}^s(\ba_1)$, which relates to the zero field discrete translation operator $\hat{T}(\ba_1)$. Note that: 
\begin{equation} \label{eq:magnetic_translation_expansion}
    \begin{split}
        \hat{t}^s(\ba_1) & = \left[ e^{-i\bq_\phi\cdot\br}\hat{T}(\ba_1) \right]^{s-2} \left[ e^{-i\bq_\phi\cdot\br}\hat{T}(\ba_1) \right]\left[ e^{-i\bq_\phi\cdot\br}\hat{T}(\ba_1) \right] \\
        & =  \left[ e^{-i\bq_\phi\cdot\br}\hat{T}(\ba_1) \right]^{s-2} e^{i \bq_\phi\cdot\ba_1} e^{-i2\bq_\phi\cdot\br} \hat{T}^2(\ba_1)\\
        & = \left[ e^{-i\bq_\phi\cdot\br}\hat{T}(\ba_1) \right]^{s-3} e^{i (1+2)\bq_\phi\cdot\ba_1} e^{-i3\bq_\phi\cdot\br} \hat{T}^3(\ba_1) \\
        & =  e^{i [1+2+\dots+(s-1)]\bq_\phi\cdot\ba_1} e^{-is\bq_\phi\cdot\br} \hat{T}^s(\ba_1)\\
        & = e^{i \frac{s(s-1)}{2}\bq_\phi\cdot\ba_1} e^{-is\bq_\phi\cdot\br} \hat{T}^s(\ba_1).
    \end{split}
\end{equation}

We also show the MTG basis states, $\ket{\Psi_{n,r}(\bk)}$ expressed using LL wavefunctions, form a complete set for $r=0,\dots p-1$. It is sufficient to show that: 
\begin{equation}
    \begin{split}
        \ket{\Psi_{n,r+p }(\bk)} & = e^{i2\pi \left( k_1 - (k_2+\frac{r}{q}) \frac{a_{1y}}{a_2}\right)}\ket{\Psi_{n,r }(\bk)},
    \end{split}
\end{equation}
namely that states at $r+p$ and $r$ differ by a complex phase factor. 
Note that by definition in Eq.~(\ref{eq:mtg_ll}),
\begin{equation} \label{eq:appd_ll_period}
    \ket{\Psi_{n,r+p }(\bk)} = \frac{1}{\sqrt{N}}\sum_{s=-\infty}^{\infty} e^{i2\pi k_1 s} \hat{t}^s(\ba_1) \ket{\psi_n\left( \frac{2\pi}{a_2}(k_2 + \frac{r}{q}+\frac{p}{q})\right)},
\end{equation}
and the LL wavefunction is expressed in Eq.~(\ref{eq:ll_wavefunction}) as: 
\begin{equation} \label{eq:ll_wavefunction_expanded}
\ket{\psi_n\left( \tilde{k}_y +\frac{2\pi}{a_2}\frac{p}{q})\right)} = e^{i\tilde{k}_y y} e^{i\frac{2\pi}{a_2}\frac{p}{q}y}\hat{T}(-\frac{2\pi}{a_2}\frac{p}{q}\ell^2\mathbf{e}_x)\hat{T}(-\tilde{k}_y\ell^2\mathbf{e}_x)\ket{n}.
\end{equation}
Here we have defined $\tilde{k}_y \equiv \frac{2\pi}{a_2}\left(k_2+\frac{r}{q}\right)$ for notational convenience. Note that 
\begin{equation}
    \frac{2\pi}{a_2}\frac{p}{q} \ell^2 \mathbf{e}_x = \frac{1}{a_2}\cdot \frac{a_{1x}a_2}{\ell^2}\cdot \ell^2 \mathbf{e}_x = a_{1x}\mathbf{e}_{x},
\end{equation}
and that 
\begin{equation}
    \hat{T}(-a_{1x}\mathbf{e}_{x}) \ket{n} = \hat{T}(-\ba_1) \ket{n}.
\end{equation}
Making use of the identity:
\begin{equation}
\hat{t}(-\ba_1) \equiv \hat{t}^\dagger(\ba_1) = \hat{T}^\dagger(\ba_1) e^{i\bq_\phi\cdot\br} = e^{i\frac{2\pi}{a_2}\frac{p}{q}a_{1y}}e^{i\frac{2\pi}{a_2}\frac{p}{q}y}\hat{T}(-\ba_1),
\end{equation}
we can rewrite Eq.~(\ref{eq:ll_wavefunction_expanded}) as: 
\begin{equation}
\ket{\psi_n\left( \tilde{k}_y +\frac{2\pi}{a_2}\frac{p}{q})\right)} = e^{-i\frac{2\pi}{a_2}\frac{p}{q}a_{1y}}e^{i\tilde{k}_y y} \hat{t}(-\ba_1)\hat{T}(-\tilde{k}_y\ell^2\mathbf{e}_x)\ket{n} = e^{-i \left(\tilde{k}_y+\frac{2\pi}{a_2}\frac{p}{q}\right) a_{1y}} \hat{t}(-\ba_1) \ket{\psi_n (\tilde{k}_y)}.
\end{equation}
Here for the last equality we used $e^{i\tilde{k}_yy}\hat{t}(-\ba_1)= e^{-i\tilde{k}_ya_{1y}}\hat{t}(-\ba_1) e^{i\tilde{k}_yy} $. Substituting this back into Eq.~(\ref{eq:appd_ll_period}), 
\begin{equation}
    \begin{split}
        \ket{\Psi_{n,r+p }(\bk)} & = e^{-i\frac{2\pi}{a_2}\left(k_2 + \frac{r+p}{q} \right)a_{1y}}\frac{1}{\sqrt{N}}\sum_{s=-\infty}^{\infty} e^{i2\pi k_1 s} \hat{t}^s(\ba_1) \hat{t}(-\ba_1)\ket{\psi_n\left( \frac{2\pi}{a_2}(k_2 + \frac{r}{q})\right)}\\
        & = e^{i2\pi \left( k_1- (k_2 + \frac{r+p}{q} )\frac{a_{1y}}{a_2}\right)}\frac{1}{\sqrt{N}}\sum_{s=-\infty}^{\infty} e^{i2\pi k_1 (s-1)} \hat{t}^{s-1}(\ba_1) \ket{\psi_n\left( \frac{2\pi}{a_2}(k_2 + \frac{r}{q})\right)}\\
        & = e^{i2\pi \left( k_1- (k_2 + \frac{r+p}{q} )\frac{a_{1y}}{a_2}\right)} \ket{\Psi_{n,r}(\bk)}.
    \end{split}
\end{equation}

\section{Overlap matrix}
Here we show how to calculate the overlap matrix between MTG basis states generated from zero field hybrid Wannier states, Eq.~(\ref{eq:hWS_overlap}). Specifically, we define the overlap matrix: $\Lambda_{\alpha r_1,\beta r_2}(k_1,k_2)$ where $k_1\in[0,1)$, $k_2\in[0,1/q)$, and $r_{1,2}=0,\dots q-1$. It can be calculated by switching to the Bloch basis, 
\begin{equation}
    \begin{split}
        \Lambda_{\alpha r_1,\beta r_2}(k_1,k_2) &= \sum_s e^{i2\pi k_1 s}e^{i\frac{s(s-1)}{2}\bq_\phi\cdot\ba_1} \bra{w_\alpha(0,k_2+\frac{r_1}{q})}e^{-is\bq_\phi\cdot \br}\ket{w_\beta(s,k_2+\frac{r_2}{q})}\\
        & = \sum_s e^{i2\pi k_1 s}e^{i\frac{s(s-1)}{2}\bq_\phi\cdot\ba_1} \frac{1}{N_1} \sum_{\bar{k}_1\bar{p}_1 } e^{-i2\pi \bar{p}_1 s}\bra{\psi_\alpha(\bar{k}_1,k_2+\frac{r_1}{q})}e^{-is\bq_\phi\cdot \br}\ket{\psi_\beta(\bar{p}_1,k_2+\frac{r_2}{q})}.
    \end{split}
\end{equation}
On the second line we introduced the Bloch states $\ket{\psi_\alpha(\bk)} = \sum_{n} U_{n,\alpha}(\bk)\ket{\psi_{n}(\bk)}$, with the distinction that the subscript $\alpha$ labels Fourier transforms of the hWS, whereas $n$ labels Bloch eigenstates. Second line is calculated in the plain wave basis, by writing down $\bra{\br}\ket{\psi_\alpha(\bk)}=\sum_{\bg} e^{i(\bk+\bg)\cdot\br}u_{\bg}(\alpha\bk)$. We arrive at the following expression: 
\begin{equation}
\begin{split}
     \Lambda_{\alpha r_1,\beta r_2}(k_1,k_2) &=  \sum_s e^{i2\pi k_1 s}e^{i\frac{s(s-1)}{2}\bq_\phi\cdot\ba_1} \frac{1}{N_1} \sum_{\bar{k}_1\bar{p}_1 }\sum_{\bg,\bg'} e^{-i2\pi \bar{p}_1 s}u^*_\bg(\alpha,\bar{k}_1,k_2+\frac{r_1}{q})u_{\bg'}(\beta,\bar{p}_1,k_2+\frac{r_2}{q})  \\
     & \times \frac{1}{\mathcal{A}}\int\mathrm{d}^2\br e^{-i(\bar{\bk}+\bg)\cdot\br}e^{-is\bq_\phi\cdot \br}e^{i(\bar{\bp}+\bg')\cdot\br}.
\end{split}
\end{equation}
The second line gives the constraint $\bar{\bk}+\bg+s\bq_\phi=\bar{\bp}+\bg'$, which is equivalent to: 
\begin{equation}
    \bar{k}_1 + l_1 + \frac{sp}{q}\frac{a_{1y}}{a_2} = \bar{p}_1 + l_1',\ \frac{r_1}{q}+ l_2 + \frac{sp}{q} = \frac{r_2}{q}+ l_2'.
\end{equation}
\section{Real space integration of $x$ and $x^2$} 
The single-electron Hamiltonian in a finite magnetic field contains terms such as $x\hat{p}_y$ and $x^2$, we hereby give analytical expressions for their real space integrations. In evaluating the matrix elements for the Hamiltonian, boundary terms do not matter, as the exponential localization of the hWSs guarantees that the main contributing terms to $\sum_{s}$ in Eq.~(\ref{eq:Hmat_hws}) are restricted to the vicinity of $s=0$. We therefore choose the real space domain to contain $N_1$ and $N_2$ unit cells in the $\ba_1$ and $\ba_2$ directions, and place the origin of the coordinate system to be at the center of the real space domain. Specifically the integral is performed in the following manner:
\begin{equation}
    \frac{1}{\mathcal{A}}\int \mathrm{d}^2\br [...]= \frac{1}{N_1N_2}\frac{1}{a_{1x}a_2}\int_{-\frac{N_1}{2}a_{1x}}^{\frac{N_1}{2}a_{1x}} \mathrm{d}x\int_{\frac{a_{1y}}{a_{1x}}x-\frac{N_2}{2}a_2}^{\frac{a_{1y}}{a_{1x}}x+\frac{N_2}{2}a_2} \mathrm{d}y [...].
\end{equation}

As a result, 
\begin{equation}
    \frac{1}{\mathcal{A}}\int \mathrm{d}^2\br \left( \frac{x}{a_{1x}} \right)e^{i\bq\cdot\br} = \delta_{q_y,0} (-i)\frac{\tilde{q}_{x}\cos(\frac{N_1}{2}\tilde{q}_x)-\frac{2}{N_1}\sin(\frac{N_1}{2}\tilde{q}_x)}{\tilde{q}_x^2}.
\end{equation}
and
\begin{equation}
    \frac{1}{\mathcal{A}}\int \mathrm{d}^2\br \left( \frac{x}{a_{1x}} \right)^2e^{i\bq\cdot\br} = \delta_{q_y,0} \frac{4\tilde{q}_x\cos(\frac{N_1}{2}\tilde{q}_x)+(-\frac{8}{N_1}+N_1\tilde{q}_x^2)\sin(\frac{N_1}{2}\tilde{q}_x)}{2\tilde{q}_x^3}.
\end{equation}
where we have defined $\tilde{q}_x\equiv q_xa_{1x}$.

\section{Overlap between MTG basis states generated from LL approach and hWS approach}
To demonstrate the accuracy of the hWS approach we need to calculate its overlap with the exact wavefunctions which can be obtained using LL approach, as discussed in Eq.~(\ref{eq:hws_ll}). Here we calculate the following expression: 
\begin{equation}
    \Lambda_{\Psi,W}(n r_1,\alpha r_2;\bk) \equiv \bra{\Psi_{n,r_1}(\bk)}\ket{W_{\alpha,r_2}(\bk)},
\end{equation}
where $n$ runs over LL indices, $\alpha$ runs over hWS indices, $r_1=0\dots,p-1$, and $r_2=0,\dots q-1$. Using the definitions Eqs.~(\ref{eq:mtg_ll}) and (\ref{eq:mtg_hw}), we have: 
\begin{equation}
\begin{split}
      \Lambda_{\Psi,W}(n r_1,\alpha r_2;\bk) & = \sum_{s} e^{-i2\pi k_1 s}\bra{\psi_n(k_2+\frac{r_1}{q})} \hat{t}_{\ba_1}^{-s}\ket{w_{\alpha}(0,k_2+\frac{r_2}{q})}\\
      & =\sum_{s} e^{-i2\pi k_1 s}e^{-i\frac{s(s-1)}{2}\bq_\phi\cdot\ba_1}e^{i2\pi(k_2+\frac{r_1}{q})\frac{a_{1y}}{a_2}s}\frac{1}{\sqrt{N_1}}\sum_{\bar{k}_1}\bra{\psi_n(k_2+\frac{r_1}{q}-\frac{sp}{q})} \ket{\psi_{\alpha}(\bar{k}_1,k_2+\frac{r_2}{q})}.
\end{split}
\end{equation}
We evaluate the overlap between a LL wavefunction with a Bloch wavefunction as follows: 
\begin{equation}
\begin{split}
      & \bra{\psi_n(k_2+\frac{r_1}{q}-\frac{sp}{q})} \ket{\psi_{\alpha}(\bar{k}_1,k_2+\frac{r_2}{q})} \\
      = &\frac{1}{N_{2}a_2}\frac{1}{\sqrt{N_1a_{1x}\ell}}\sum_{\bg} u_{\alpha\bg} \int \mathrm{d}^2\br \phi_n\left( \frac{x}{\ell} +\frac{2\pi}{a_2}(k_2+\frac{r_1-sp}{q})\ell\right) e^{-i\frac{2\pi}{a_2}(k_2+\frac{r_1-sp}{q})y} e^{i(\bar{\bk}+\bg)\cdot\br}\\
      = & \frac{1}{\sqrt{N_1a_{1x}\ell}}\sum_{\bg} u_{\alpha\bg} \delta_{\frac{r_1-sp}{q},\frac{r_2}{q}+l_2}\int \mathrm{d}x \phi_n\left( \frac{x}{\ell} +(\bar{k}_y+g_y)\ell\right)  e^{i(\bar{k}_x+g_x)x}\\
     = & \frac{1}{\sqrt{N_1a_{1x}\ell}}\sum_{\bg} u_{\alpha\bg}\delta_{\frac{r_1-sp}{q},\frac{r_2}{q}+l_2} e^{-i(\bar{k}_x+g_x)(\bar{k}_y+g_y)\ell^2} \int \mathrm{d}x \phi_n\left(\frac{x}{\ell}\right) e^{i(\bar{k}_x+g_x)x} \\
     = & \frac{1}{\sqrt{N_1}}\sqrt{\frac{\ell}{a_{1x}}}\sum_{\bg} u_{\alpha\bg}\delta_{\frac{r_1-sp}{q},\frac{r_2}{q}+l_2} e^{-i(\bar{k}_x+g_x)(\bar{k}_y+g_y)\ell^2} \sqrt{2\pi} (i)^n \phi_n\left((\bar{k}_x+g_x)\ell\right)
\end{split}
\end{equation}
Here we have defined $\bar{\bk}=\bar{k}_1\bg_1+(k_2+\frac{r_2}{q})\bg_2\equiv \bar{k}_x\mathbf{e}_x+\bar{k}_y\mathbf{e}_y$. On the second line we used the following definitions for the Bloch and LL wavefunctions:
\begin{align}
    \bra{\br}\ket{\psi_\alpha(\bk)} &= \frac{1}{\sqrt{N_2a_2}}\frac{1}{\sqrt{N_1a_{1x}}} \sum_{\bg} u_{\alpha\bg}(\bk) e^{i(\bk+\bg)\cdot\br},\\
    \bra{\br}\ket{\psi_{n}(k_2)} &= \frac{1}{\sqrt{N_2a_2}} \frac{1}{\sqrt{\ell}}\phi_n\left( \frac{x}{\ell}+\frac{2\pi}{a_2}k_2\ell \right)e^{i\frac{2\pi}{a_2}k_2y},\\
\end{align}
where: 
\begin{equation} \label{eq:harmonic_oscillator}
    \phi_n\left( \frac{x}{\ell}\right) = \frac{1}{\pi^{1/4}\sqrt{2^nn!}}e^{-\frac{x^2}{2\ell^2}}H_n\left(\frac{x}{\ell}\right).
\end{equation}

\section{Tight-binding model and Peierls substitution}
Here we present a calculation of the Hofstadter spectrum for a tight-binding model on a square lattice with one orbital per site and longer ranged hoppings. The spectra is then compared to both the exact LL approach and the hybrid Wannier approach discussed in the main text. 

The tight binding Hamiltonian in zero magnetic field is given by: 
\begin{equation}
    \hat{H}_{TB} = \sum_{\br_1,\br_2} t(\br_1-\br_2) c^{\dagger}_{\br_1}c_{\br_2} + h.c.,
\end{equation}
where $\br_i = m_i \ba_1 + n_i\ba_2$ labels positions of lattice sites, with $\{m_i,n_i\}\in \mathbb{Z}$. $t(\br_1-\br_2)$ is the hopping amplitude between the two sites, and only dependent on the relative coordinate due to discrete translation symmetry $\hat{T}(\ba_i),i=1,2$. Its Fourier transform is the energy dispersion: 
\begin{equation}
    \varepsilon_{\bk} =\sum_{\br_1-\br_2}t(\br_1-\br_2)e^{-i\bk\cdot(\br_1-\br_2)},
\end{equation}
where $\bk=k_1\bg_1+k_2\bg_2$ is the wavevector in the  Brillouin zone, with $k_1,k_2\in[0,1)$. $N_1$ and $N_2$ are number of lattice sites along the $\ba_1$ and $\ba_2$ directions respectively. 

In a finite magnetic field, Peierls substitution \cite{Peierls1933} attaches a phase to the hopping amplitude: 
\begin{equation} \label{eq:peierls_hopping}
    t(\br_1-\br_2) e^{-i\frac{e}{\hbar c}\int_{\br_1}^{\br_2} \mathrm{d}\br'\cdot \bA(\br')} = t(\br_1-\br_2) e^{-\frac{i}{2\ell^2}(r_{2y}-r_{1y})(r_{2x}+r_{1x})},
\end{equation}
where on the right hand side of the equation we used the Landau gauge $\bA=B x\mathbf{e}_y$.

At rational magnetic flux ratios $\phi/\phi_0=p/q$, the phase factor in Eq.~(\ref{eq:peierls_hopping}) reduces to: 
\begin{equation}
    -\frac{2\pi p}{q} (n_2-n_1)\frac{m_1+m_2}{2},\ \br_{i=1,2}\equiv m_i\ba_1+n_i\ba_2,\ {m_i,n_i}\in \mathbb{Z}.
\end{equation}

Therefore, the finite field Hamiltonian is invariant under discrete translations $\hat{T}(q\ba_1)$ and $\hat{T}(\ba_2)$. As a result, the magnetic unit cell is enlarged along the $\ba_1$ direction by $q$ times. We relabel the lattice sites as: 
\begin{equation}
    \br_1\rightarrow \bR_1+\vec{\tau}_{\alpha},\ \br_2\rightarrow \bR_2+\vec{\tau}_{\beta},
\end{equation}
where $\bR_{i=1,2}\equiv m_i(q\ba_1)+n_i(\ba_2)$, and $\vec{\tau}_{\alpha}\equiv \alpha \ba_1,\ \alpha= 0,\dots q-1$. The Peierls substituted hopping amplitude is therefore rewritten as:
\begin{equation}
    t(\bR_1+\vec{\tau}_\alpha-\bR_2-\vec{\tau}_\beta) e^{-i {2\pi p}(n_2-n_1)\left(\frac{m_1+m_2}{2}+\frac{\alpha+\beta}{2q} \right)}.
\end{equation}

We apply discrete Fourier transformation for the fermion annihilation operator: 
\begin{equation}
    c_{\bR_1+\vec{\tau}_\alpha} = \frac{1}{\sqrt{N_1N_2/q}}\sum_{k_1\in[0,1/q)}\sum_{k_2\in[0,1)} c_{\alpha,k_1,k_2}e^{i2\pi k_1 (q m_1+\alpha)}e^{i2\pi k_2 n_1}.
\end{equation}
It is straightforward to show that the finite field Hamiltonian is diagonal with respect to quantum numbers $\{k_1,k_2\}$, and: 
\begin{equation} \label{eq:tb_peierls}
\begin{split}
    \hat{H}_{TB}(\bB) & = \frac{q}{N_1N_2}\sum_{\alpha,\beta}\sum_{k_1,k_2} c^{\dagger}_{\alpha,k_1,k_2}c_{\beta,k_1,k_2}\frac{1}{N_1N_2}\sum_{p_1,p_2} \varepsilon_{p_1,p_2} \\
    \times & \sum_{m_1,n_1,m_2,n_2}e^{i2\pi p_2(n_1-n_2)}e^{i2\pi p_1\left[(m_1-m_2)q+(\alpha-\beta)\right]}\\
    \times &  e^{-i2\pi p(n_2-n_1)\left(\frac{m_1+m_2}{2}+\frac{\alpha+\beta}{2q} \right)}e^{-i2\pi k_2(n_1-n_2)}e^{-i2\pi k_1\left[(m_1-m_2)q+(\alpha-\beta)\right]}.
\end{split}
\end{equation}
Here $\bp=p_1\bg_1+p_2\bg_2$ is defined in the zero field Brillouin zone with $p_1,p_2\in[0,1)\times[0,1)$. We define center of mass and relative coordinates: 
\begin{equation}
    M= \frac{m_1+m_2}{2},\ m=m_1-m_2,\ N= \frac{n_1+n_2}{2},\ n=n_1-n_2.
\end{equation}

Summing over center of mass position $Mq\ba_1+N\ba_2$ trivially leads to:
\begin{equation} 
\begin{split}
    \hat{H}_{TB}(\bB) & = \frac{1}{N_1N_2}\sum_{\alpha,\beta}\sum_{k_1,k_2} c^{\dagger}_{\alpha,k_1,k_2}c_{\beta,k_1,k_2}\sum_{p_1,p_2} \varepsilon_{p_1,p_2} \\
    \times & \sum_{m,n} e^{i2\pi (p_2-k_2-\frac{p(\alpha+\beta)}{2q})n}e^{i2\pi (p_1-k_1)\left[mq+(\alpha-\beta)\right]}.
\end{split}
\end{equation}

Summing over $n$ and $m$ leads to: 
\begin{align} 
    \hat{H}_{TB}(\bB) & \equiv \sum_{\alpha,\beta}\sum_{k_1,k_2} c^{\dagger}_{\alpha,k_1,k_2}\hat{T}_{\alpha,\beta}(k_1,k_2)c_{\beta,k_1,k_2},\\
    \hat{T}_{\alpha,\beta}(k_1,k_2) & \equiv \frac{1}{q}\sum_{r_1,p_2} \varepsilon_{k_1+\frac{r_1}{q},p_2} e^{i2\pi \frac{r_1}{q}(\alpha-\beta)}\delta_{p_2,[k_2+\frac{p(\alpha+\beta)}{2q}]_1}.
\end{align}

On the second line we have defined $p_1=[p_1]_{1/q}+\frac{r_1}{q}$ where $r_1=0,\dots q$, and that $k_1=[p_1]_{1/q}$. At any given wavevector $\bk\equiv k_1\bg_1+k_2\bg_2$, $\hat{T}_{\alpha,\beta}(k_1,k_2)$ is a $q\times q$ matrix. Diagonalizing $\hat{T}_{\alpha,\beta}(k_1,k_2)$ gives the $q$ Hofstadter subbands for a given flux ratio $\phi/\phi_0=p/q$. 

\begin{figure}
    \centering
    \includegraphics[width=\linewidth]{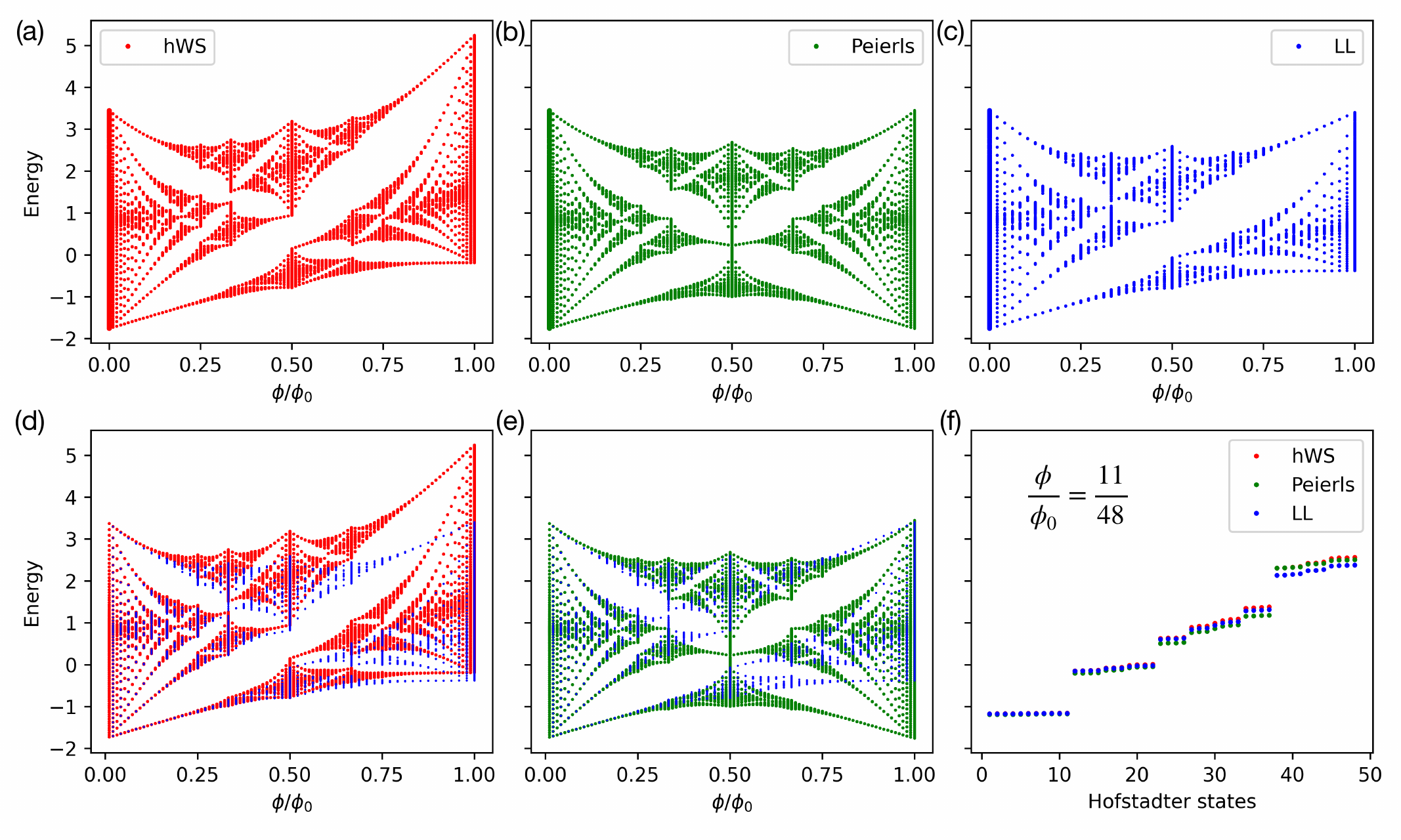}
    \caption{Hofstadter spectrum for square lattice potential used in Fig.~\ref{fig:hofstadter_comparison}(a-c) of the main text, calculated using hybrid Wannier (a), Peierls substitution (b), and Landau level (c) methods. (d,e) are more detailed overlay comparisons. (f) is the comparison of the spectra calculated at wavevector $k_1=k_2=0$ at magnetic flux ratio $\phi/\phi_0=11/48$.}
    \label{fig:hofstadter_comparison_appendix}
\end{figure}

In Fig.~\ref{fig:hofstadter_comparison_appendix} we compare the Hofstadter spectra computed using the above Peierls substitution method (green) to the Landau level method (blue) and the hybrid Wannier method (red). The Peierls substitution method is in reasonably good agreements with the two other approaches at lower magnetic flux ratios [Fig.~\ref{fig:hofstadter_comparison_appendix}(f)]. However at higher fluxes, the method leads to qualitatively different spectra. For example, at $\phi/\phi_0=1/2$, the Peierls substitution shows a gapless spectra at half filling, instead of gapped as shown by both Landau level and hybrid Wannier methods. Moreover, the Peierls substitution shows a periodic spectra when magnetic flux is increased by unit flux quantum, due to omitting the energetic effects of magnetic fields, i.e., $\hat{H}_\bB$ in Sec.~\ref{sec:HB}. of the main text. 

\end{widetext}

\end{document}